\documentclass[a4paper,longbibliography,twocolumn,11pt]{quantumarticle}
\pdfoutput=1
\usepackage[utf8]{inputenc}
\usepackage[english]{babel}
\usepackage[T1]{fontenc}
\usepackage{hyperref}

\usepackage{tikz}
\usepackage{lipsum}

\usepackage{amsmath} 
\usepackage{ae}
\usepackage{siunitx}
\usepackage{tikz}
\usetikzlibrary{quantikz}
\usepackage{subcaption}
\usepackage{braket}
\usepackage{amsthm}
\usepackage{amssymb}

\begin{document}

\title{Supervised learning of random quantum circuits via scalable neural networks}
\author{Simone Cantori}
\affiliation{School of Science and Technology, Physics Division, Universit{\`a}  di Camerino, 62032 Camerino (MC), Italy}
\orcid{0000-0002-6071-9987}
\author{David Vitali}
\affiliation{School of Science and Technology, Physics Division, Universit{\`a}  di Camerino, 62032 Camerino (MC), Italy}
\affiliation{INFN-Sezione di Perugia, 06123 Perugia, Italy}
\affiliation{CNR-INO, 50125 Firenze, Italy}
\orcid{0000-0002-1409-7136}
\author{Sebastiano Pilati}
\affiliation{School of Science and Technology, Physics Division, Universit{\`a}  di Camerino, 62032
Camerino (MC), Italy}
\affiliation{INFN-Sezione di Perugia, 06123 Perugia, Italy}
\orcid{0000-0002-4845-6299}

\maketitle

\begin{abstract}
Predicting the output of quantum circuits is a hard computational task that plays a pivotal role in the development of universal quantum computers.
Here we investigate the supervised learning of output expectation values of random quantum circuits.
Deep convolutional neural networks (CNNs) are trained to predict single-qubit and two-qubit expectation values using databases of classically simulated circuits. These circuits are represented via an appropriately designed one-hot encoding of the constituent gates.
The prediction accuracy for previously unseen circuits is analyzed, also making comparisons with small-scale quantum computers available from the free IBM Quantum program. The CNNs often outperform the quantum devices, depending on the circuit depth, on the network depth, and on the training set size.
Notably, our CNNs are designed to be scalable. This allows us exploiting transfer learning and performing extrapolations to circuits larger than those included in the training set. 
These CNNs also demonstrate remarkable resilience against noise, namely, they remain accurate even when trained on (simulated) expectation values averaged over very few measurements.
\end{abstract}

\maketitle

\section{Introduction}
Universal quantum computers  promise to solve some relevant computational problems which are intractable for classical computers~\cite{Steane_1998,doi:10.1142/9789812815569_0007}.
In fact, the claim of quantum speed-up is justified only when the targeted computational task  cannot be completed by any classical algorithm in a comparable computation time~\cite{troyerdefining}. In this context, machine learning techniques represent a relevant benchmark. They are alternative to direct classical simulations of quantum-circuits, which are plagued by an exponentially scaling computational cost.
In particular, supervised learning from classically simulated datasets has emerged as a promising and computationally feasible strategy to predict the ground-state properties of complex quantum systems, including, e.g., small molecules~\cite{faber2017prediction,doi:10.1021/acs.jpclett.5b00831}, solid-state systems~\cite{PhysRevB.89.205118,RYCZKO2018134}, atomic gases~\cite{ML_Q4,10.21468/SciPostPhys.10.3.073}, and protein-ligand complexes~\cite{ballester2010machine,KHAMIS2015135}.
Interestingly, it has recently been proven that data-based algorithms can solve otherwise classically intractable computational tasks~\cite{huang2021power}, including predicting ground-state properties of quantum systems, and rigorous guarantees on the accuracy and on the scaling of the required training set size have been demonstrated~\cite{ML_Q3}.
%
%
Still, producing training sets for supervised learning via classical computers quickly becomes unfeasible as the system size increases. In the context of ground-state simulations, this problem has been addressed via scalable neural networks~\cite{mills2019extensive,ML_Q5}. These allow performing  transfer learning from small to large systems~\cite{10.21468/SciPostPhys.10.3.073, jungsize}, and even to extrapolate to sizes larger than those included in the training set. 
This paves the way to addressing classically intractable quantum systems.

The above considerations led us to investigate the supervised learning of gate-based quantum computers. 
Our goal is to demonstrate that deep convolutional neural networks (CNNs) can be trained to predict relevant output properties of quantum circuits, both from exact classical simulations of expectation values, as well as from noisy measurements.
Remarkably, we show that the CNNs trained on random circuits are able to emulate a broad category of quantum circuits,  including, e.g., the Berstein-Vazirani (BV) algorithm. Furthermore, thanks to an appropriately designed scalable structure, they provide accurate extrapolations for circuits larger than those included in the training set.
These findings motivate the use of quantum devices as platforms for training CNNs to emulate classically intractable circuits.

In detail, in this article we consider large ensembles on quantum circuits built with gates randomly selected from an approximate universal set.
It is worth mentioning that the sampling from random circuits is the computational task 
considered in the recent demonstrations of quantum supremacy~\cite{Google, ChinaQS}.
The target of the supervised learning is a partial description of the circuit output, namely, single-qubit and two-qubit expectation values. As we demonstrate, this limited information is still sufficient to emulate the category of quantum circuits with only one or two possible output bit strings. Interestingly, this category includes relevant circuits such as the BV algorithm~\cite{BV}.
An appropriate one-hot encoding of the constituent gates is designed to unequivocally describe the circuits. Their output is classically computed using the Qiskit library~\cite{Qiskit}, considering numbers of qubits up to $N\simeq 10$ and circuit depths (number of gates per qubit) up to $P\simeq 10$. Considerably larger circuits are also considered in the testing processes performed via extrapolation procedures.
Deep CNNs are trained to map the circuit descriptors to the output expectation values. The CNNs are tested on previously unseen circuits, and we analyse how the predictions accuracy varies as a functions of the circuit size, of the network depth, and of the training set size.
We also compare the accuracy of the trained CNNs against the one of various small quantum computers available via IBM Quantum Experience~\cite{IBMQ}. 
Generally, the CNNs outperform the freely available noisy intermediate-scale quantum (NISQ) processors, unless the circuit's depth is increased at fixed CNN parameters.
Notably, our CNNs are designed to be scalable. This allows us investigating transfer-learning and extrapolation protocols.
Specifically, we show that the learning of large circuits can be accelerated via a pretraining performed on smaller circuits. Furthermore, we employ CNNs trained on small circuits to predict the output of circuits with up to twice as many qubits.
Interestingly, CNNs also learn (from random circuits) to emulate the BV algorithm, even when the number of qubits is increased by several orders of magnitude.
Finally, we consider the training on noisy expectation values, obtained as averages over a variable number of (simulated) measurements. We find that the CNNs are able to filter this noise, providing remarkably accurate estimates of the exact expectation values even when very few measurements are considered in the training data.


The rest of the article is organized as follows: in Section~\ref{Sec2} we describe the random circuits, the one-hot encoding we design for supervised learning, the CNNs we adopt, and the target expectation values. Furthermore, we discuss the class of random circuits that can be emulated from the target values we address. The predictions of the trained CNNs are analyzed in Section~\ref{Sec4}. In the same Section these predictions are also compared with small quantum computers available from IBM Quantum Experience. Then, transfer learning and extrapolations protocols are analyzed, and we finally discuss the training on noisy expectation values.
Section~\ref{Sec5} reports our conclusions and an outlook on future perspectives.

\section{Methods}\label{Sec2}

\subsection{Representation of random circuits}
\label{representationsubsec}
Our goal is to train deep CNNs to map univocal representations of random circuits to their output expectation values.
Specifically, we consider circuits built with two single-qubit gates, namely, the T-gate ($T$) and the Hadamard gate ($H$), together with one two-qubit gate, namely, the controlled-not gate ($CX$). Notably, the set $\mathcal{S}=\{T,H,CX\}$ constitutes an approximately universal set~\cite{BOYKIN2000101,toffalori2005teoria}, meaning that any unitary operator can be implemented using these three gates.
It is worth noticing that the identity $I=\begin{bmatrix}
                        	1 & 0 \\
                        	0 & 1
                        \end{bmatrix}$ can be represented as $I=H^2$. 
Below, an extended set explicitly including the gate $I$ will be discussed.
We adopt the standard computational basis corresponding to the eigenstates of the Pauli matrix $Z=\begin{bmatrix}
                        	1 & 0 \\
                        	0 & -1
                        \end{bmatrix}$. 
In this basis, the three considered gates  are represented by the following matrices:
\begin{eqnarray}\label{Gates}
    T&=&\begin{bmatrix}
			1 & 0 \\
			0 & e^{i\frac{\pi}{4}} 	
		\end{bmatrix} \, \,
	H=\frac{1}{\sqrt{2}}\begin{bmatrix}
		1 & 1 \\
		1 & -1 	
	\end{bmatrix} \, \,\nonumber\\
	CX&=&\begin{bmatrix}
		1 & 0 & 0 & 0\\
		0 & 1 & 0 & 0\\ 
		0 & 0 & 0 & 1\\
		0 & 0 & 1 & 0 \end{bmatrix} .
\end{eqnarray}
%
%
In the following, we consider circuits with $N$ qubits and $P$ layers of gates. Therefore, the integer $P$ corresponds to the number of gates per qubit; this parameter will be referred to also as the circuit depth. 
In every layer, each qubit is processed by a gate randomly selected from the set $\mathcal{S}$. Notice that the  two-qubit gate $CX$ acts on a control and on a target qubit.  To simplify the circuit representation, we allow only one $CX$ gate at every  layer. 
Circuits with more $CX$ gates per layer can be emulated by deeper circuits satisfying the constraint.
This constraint allows us adopting a relatively simple univocal circuit representation. It is based on the following one-hot encoding of the gate acting on each qubit: the $T$-gate corresponds to the vector $(1,0,0,0)$, the $H$-gate to $(0,1,0,0)$,  the control qubit of the $CX$-gate corresponds to $(0,0,1,0)$, while the target qubit corresponds to $(0,0,0,1)$.
This map is also represented in Fig.~\ref{OneHot}.
\begin{figure}[!h]
\centering
		\begin{quantikz}[column sep = small]
			&\qw & \gate{T} & \qw  &\arrow[r]  & \hspace{2mm} \rstick{$[1,0,0,0]$} \\
			&\qw & \gate{H} & \qw  &\arrow[r] & \hspace{2mm} \rstick{$[0,1,0,0]$} \\
			&\qw & \ctrl{1} & \qw  &\arrow[r] & \hspace{2mm} \rstick{$[0,0,1,0]$} \\
			&\qw & \targ & \qw & \qw  &\arrow[r] & \hspace{2mm} \rstick{$[0,0,0,1]$} \\
		\end{quantikz}
		\caption{Visualization of the one-hot encoding that represents the quantum gates in the set $\mathcal{S}$. The full circle and the empty circle with plus sign represent the control and the target qubit of the $CX$ gate, respectively.}
		\label{OneHot}
\end{figure}
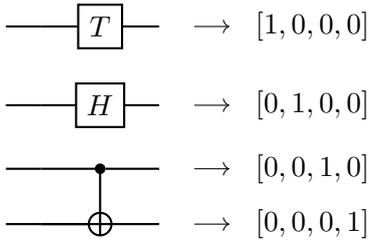
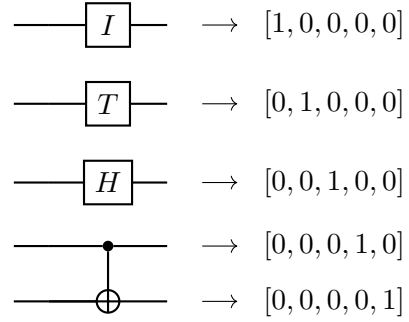
\begin{figure}[!h]
\centering
		\begin{quantikz}[column sep = small]
		    &\qw & \gate{I} & \qw  &\arrow[r]  & \hspace{2mm} \rstick{$[1,0,0,0,0]$} \\
			&\qw & \gate{T} & \qw  &\arrow[r]  & \hspace{2mm} \rstick{$[0,1,0,0,0]$} \\
			&\qw & \gate{H} & \qw  &\arrow[r] & \hspace{2mm} \rstick{$[0,0,1,0,0]$} \\
			&\qw & \ctrl{1} & \qw  &\arrow[r] & \hspace{2mm} \rstick{$[0,0,0,1,0]$} \\
			&\qw & \targ & \qw & \qw  &\arrow[r] & \hspace{2mm} \rstick{$[0,0,0,0,1]$} \\
		\end{quantikz}
		\caption{Visualization of the one-hot encoding that represents the quantum gates in the set $\mathcal{S}^*$.}		
		\label{OneHot_Identity}
\end{figure}
Therefore, the feature vector representing a random circuit is a four-channel two-dimensional binary matrix with dimensions $N\times P\times 4$. 
Analogous gate-based descriptions of quantum circuits have been adopted in Refs.~\cite{encoding,Zhang_2021}.
Despite the constraint on the number of $CX$ gates per layer, the number $Q$ of possible  circuits rapidly grows with $N$ and $P$. This number can be computed as:
\begin{equation}\label{Q}
	Q=\sum_{m=0}^{P}2^{N\cdot P-2m}\binom{P}{m}2^m\binom{N}{2}^m \, ,
\end{equation}
where $m$ is the number of $CX$-gates in the circuit. The first term, namely, $2^{N\cdot P-2m}$, represents  the possible combinations considering only the $T$-gates and the $H$-gates. The second term, namely,  $\binom{P}{m}$, corresponds to the possible combinations of the $CX$-gates in $P$ layers. The third term, namely, $2^m$, corresponds to the choice of the control and of the target qubit for each $CX$ gate. Finally, the term $\binom{N}{2}^m$ corresponds to the available pairs for each $CX$-gate. 
For example, for the smallest circuit size considered in this article, namely, $N=3$ and $P=5$, one has $Q=3.2\times 10^{6}$ possible random circuits. For the largest size, namely, $N=20$ and $P=6$, one has the astronomic number $Q\simeq10^{48}$. 
This means that it is virtually impossible to create a dataset exhausting  the whole ensemble of possible circuits. 
We instead resort to the generalization capability of deep CNNs. These are expected to provide accurate predictions for previously unseen instances, even when trained on (largely non-exhaustive) training sets including a number $N_\mathrm{train}\ll Q$ of training instances.

While the set $\mathcal{S}$ is, in principle, universal, the choice of operating on all qubits in every layer implies that some unitary operators cannot be represented. Therefore, we also consider the extended set $\mathcal{S}^*=\mathcal{S} \bigcup \{I\}$, where the identity is explicitly included. 
For this set, a five channel one-hot encoding is needed. We use the map represented in Fig.~\ref{OneHot_Identity}.
Most of the results reported in this article are based on the set $\mathcal{S}$, unless stated otherwise. Notably, this set is flexible enough to represent the BV algorithm, which we use as a relevant test bed.
For a few representative test cases, we also consider the extended gate set $\mathcal{S}^*$, finding very similar performances.

\subsection{Target values}
\begin{figure}[]
	\centering
	\includegraphics[width=\columnwidth]{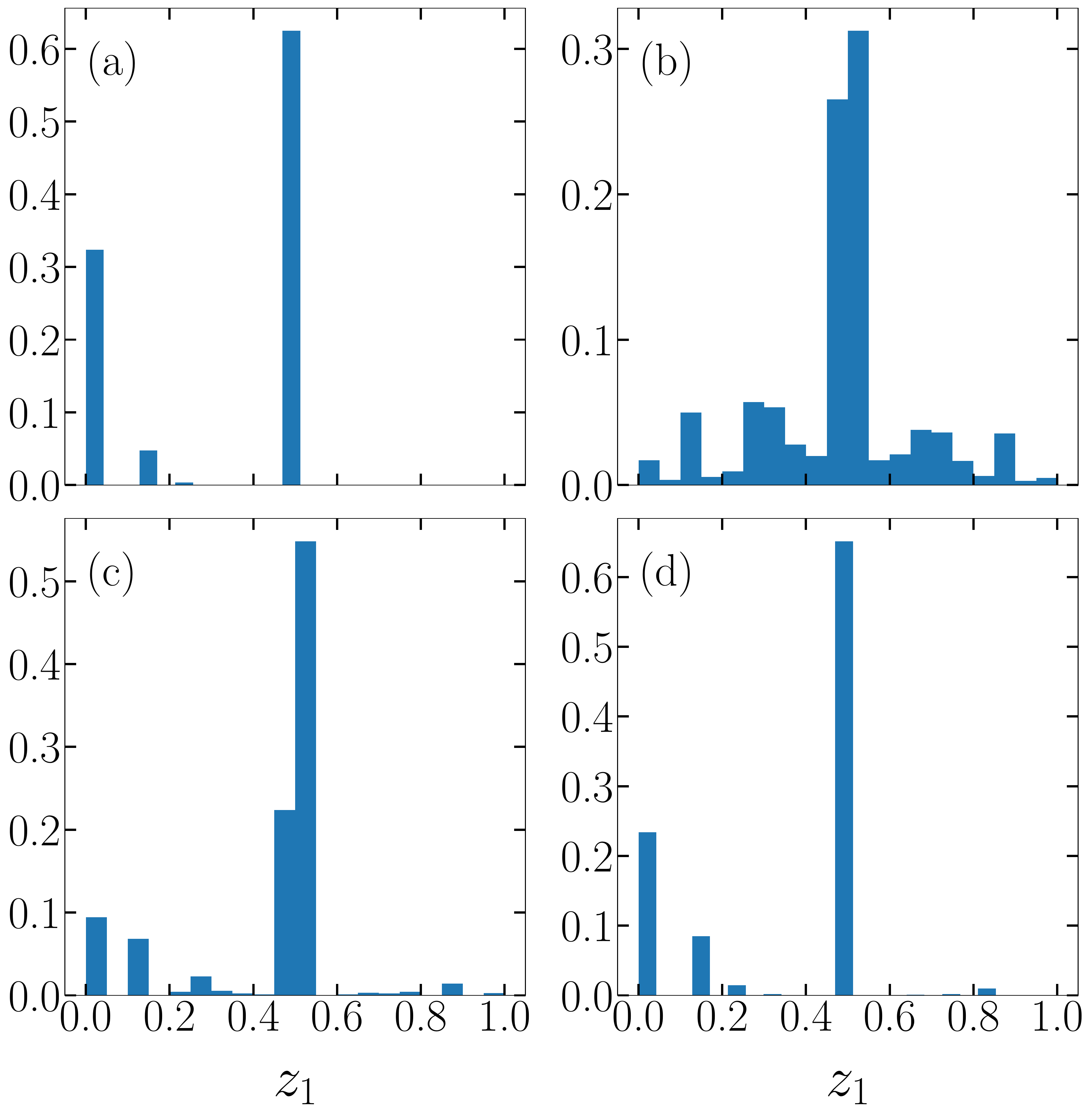}
	\caption{Normalized histograms of the rescaled single-qubit expectation values $z_1=1-\frac{\braket{Z_1} + 1}{2}$ over ensembles of random circuits.
	Different number of qubits $N$ and circuit depths $P$ are considered in the four panels: $N=3$ and $P=5$ (a); $N=3$ and $P=20$ (b); $N=5$ and $P=10$ (c); $N=20$ and $P=6$ (d). 
	The ensembles in (a), (b), and (c), and the one in (d) include 5000 and 2000 circuits, respectively.}
	\label{hists}
\end{figure}

The output state of a quantum circuit can be written as
\begin{equation}
	\ket{\psi}=U\ket{0}^{\otimes N},
\end{equation}
where the tensor product $\ket{0}^{\otimes N}\equiv \ket{0}_1\otimes \dots \ket{0}_N$ is the input state and the unitary operator $U$ represents the sequence of quantum gates that constitute the circuit. 
Here and in the following, we indicate with $\ket{0}_i$ and $\ket{1}_i$ the eigenvectors of the Pauli operator $Z_i$ acting on qubit $i=1,\dots,N$, such that $Z_i \ket{0}_i= \ket{0}_i$ and $Z_i \ket{1}_i= -\ket{1}_i$.
With this notation, each state $\ket{\textbf{x}}$ of the many-qubit computational basis  corresponds to a bit string $\textbf{x}=x_1\dots x_N$, where $x_i=0,1$ for $i=1,\dots,N$.
Our goal is to perform supervised learning of output expectation values.
First, we focus on the single-qubit expectation values
\begin{equation}
	\braket{Z_i}\equiv\braket{\psi|Z_i|\psi}.
\end{equation}
These expectation values can be computed as 
\begin{equation}
	\braket{Z_i} = \left | \braket{\psi|0}_i \right |^2 - \left | \braket{\psi|1}_i \right |^2 \, .
\end{equation}
%
%
%
It is convenient to perform the following rescaling:
\begin{equation}\label{totargetvalue}
	z_i=1-\frac{\braket{Z_i} + 1}{2},
\end{equation}
so that $z_i\in[0,1]$, and $z_i=0$ corresponds to $\ket{0}_i$, while $z_i=1$ corresponds to $\ket{1}_i$.
The rescaled variable $z_i$ is the first target value we address for supervised learning.
It is worth anticipating that we will consider both CNNs designed to predict only one expectation value, say, $z_1$, and CNNs that simultaneously predict all single-qubit expectation values $z_i$, for $i=1,\dots,N$. This is discussed with more details in subsection~\ref{cnnsubsection}.\\
For illustrative purposes, we show in Fig.~\ref{hists} the distribution of the target value $z_1$ over an ensemble of random circuits. Four representative circuit sizes are considered.
%
%
%
One notices rather pronounced peaks around $z_1=1/2$ for more qubits and/or deeper circuits.
The second target values we consider are the two-qubit  expectation values $\braket{Z_iZ_j}$, where, in general, $i,j=1,2,...,N$.
Specifically, we focus on the case $i=1$ and $j=2$, and the target value is the rescaled variable:
\begin{equation}
\label{eqz12}
	z_{12}=1-\frac{\braket{Z_1Z_2} + 1}{2} \, .
\end{equation}
Clearly, single-qubit and two-qubit Pauli-$Z$ expectation values represent a limited description of the circuit output. However, this information is sufficient to unambiguously identify the output of a significant category of quantum circuits. This category is described in the following subsection, where we also discuss some relevant examples belonging to the category.

\subsection{Emulable quantum algorithms}
\label{emulable}
%
%
For certain quantum algorithms, only a small subset of the $2^N$ output bit strings have non-zero measurement probability. In fact, some relevant circuits have only one possible outcome (in the absence of noise and errors). These are discussed below. 
First, we consider the more generic category for which only two output bit strings, which we indicate as $\textbf{a}$ and $\textbf{b}$, have non-zero measurement probabilities 
$p(\textbf{a}) = \left|\braket{ \textbf{a} | \psi } \right|^2$ and 
$p(\textbf{b}) = \left|\braket{ \textbf{b} | \psi } \right|^2$.
With this notation, one has
$p(\textbf{a})+p(\textbf{b})= 1$. 
For this category of circuits, the expectation values $\braket{Z_i}$ and $\braket{Z_iZ_j}$ provide all the information required to unambiguously identify the bit strings $\textbf{a}$ and $\textbf{b}$.
This statement is proven here by providing an explicit algorithm. It is assumed that $p(\textbf{x})=0$ for $\textbf{x} \neq \textbf{a}, \textbf{b}$ and that the expectation values mentioned above are known. 
It is worth emphasising that, in fact, if $p(\textbf{a})\neq p(\textbf{b})$, knowledge of the single-qubit expectation values suffices. 
Indeed, the values $\braket{Z_iZ_j}$ are only used when  $p(\textbf{a})= p(\textbf{b})$ --- see case iv) below --- and such case is easily identified since one must have $\braket{Z_i}=0$ for at least one qubit $i$.
%
%
	%
%
\begin{proof}
	The qubit are analyzed in the order $i=1,\dots,N$.
	Four possible cases have to be separately treated:\\ 
	i) If $\braket{Z_i}=1$, the corresponding bits are set to $a_i=b_i=0$.\\
	ii) If $\braket{Z_i}=-1$, one sets $a_i=b_i=1$.\\
	iii) If $\braket{Z_i} \in (-1,1)$ and either
	$i=1$ or
	$i>1$ with $a_j=b_j$ for $j=1,\dots,i-1$ 
	we arbitrarily set, without loss of generality, $a_i=1$ and $b_i=0$.
	Notice that we can also infer the two probabilities: $p(\textbf{a})=\frac{1-\braket{Z_i}}{2}$ and $p(\textbf{b})=\frac{1+\braket{Z_i}}{2}$.\\
	iv) Otherwise, when $p(\textbf{a}) \neq p(\textbf{b})$ (this is known from case iii), two expectation values are possible. One is $\braket{Z_i}=\braket{Z_j}$, where the integer $j\in[1,i-1]$ is the first index such that $\braket{Z_i} \in (-1,1)$ (at least one exists); in this case one sets $a_i=a_j$ and $b_i=b_j$. If, instead, $\braket{Z_i}\neq\braket{Z_j}$, one sets $a_i=1-a_j$ and $b_i=1-b_j$.
	When $p(\textbf{a})= p(\textbf{b})=1/2$, one must have $\braket{Z_i}=0$, and we also know that an integer $j\in[1,i-1]$ such that $\braket{Z_j}=0$ exists. In this situation, $\braket{Z_iZ_j}=\pm 1$. If $\braket{Z_iZ_j}=1$, one sets $a_i=a_j$ and $b_i=b_j$. If, instead, $\braket{Z_iZ_j}=-1$, one sets $a_i=1-a_j$ and $b_i=1-b_j$.
\end{proof}

As discussed in the previous paragraph, the single-qubit expectation values $\braket{Z_i}$, eventually combined with $\braket{Z_iZ_j}$, are sufficient to identify the output bit strings when only two of them have non-zero probability.
Clearly, the values $\braket{Z_i}$ are sufficient when only one bit string is possible.
Interestingly, some paradigmatic quantum algorithm belong to this group.
A relevant example is the Deutsch-Jozsa algorithm~\cite{DJ}. This allows assessing whether a boolean function $f:\{0,1\}^{(N-1)}\rightarrow \{0,1\}$  is either constant or balanced.
The algorithm involves measuring  the first $N-1$ qubits. If the outcome corresponds to the bit string $00...0$, the function is constant, otherwise the function is balanced. 
Predicting $N-1$ single-qubit expectation values  allows reaching the same result. Practically, if $\braket{Z_i}=1$  for $i=1,2,...,N-1$, the only possible bit string is the bit string $00...0$, corresponding to a constant function.
Otherwise, the function is balanced. 
Also in the BV algorithm~\cite{BV} there is  only one possible outcome.
This algorithm is designed to identify an unknown bit string $\textbf{w}\in \{0,1\}^{N-1}$, assuming we are given an oracle that implements the boolean function 
$f: \{0,1\}^{N-1}\rightarrow \{0,1\}$ defined as $f(\textbf{x})=\textbf{x}\circledcirc \textbf{w}$, where the symbol $\circledcirc$ represents the dot product modulo 2.
Notably, the BV algorithm provides the answer with one function query, outperforming classical computers which require $N-1$ queries.
Single-qubit expectation values allow identifying $\textbf{w}$: $\braket{Z_i}=1$ corresponds to $w_i=0$, while $\braket{Z_i}=-1$ corresponds to $w_i=1$.
The Grover algorithm with two searched items~\cite{Grover} and the quantum counting algorithm~\cite{Counting} have two output bit strings with much higher probabilities than the other strings. Therefore one could use the predictions for $\braket{Z_i}$ and $\braket{Z_iZ_j}$ to emulate also these latter algorithms.

\subsection{Convolutional neural networks and training protocol}
\label{cnnsubsection}
\begin{figure}[]
	\centering
	\includegraphics[width=0.92\columnwidth]{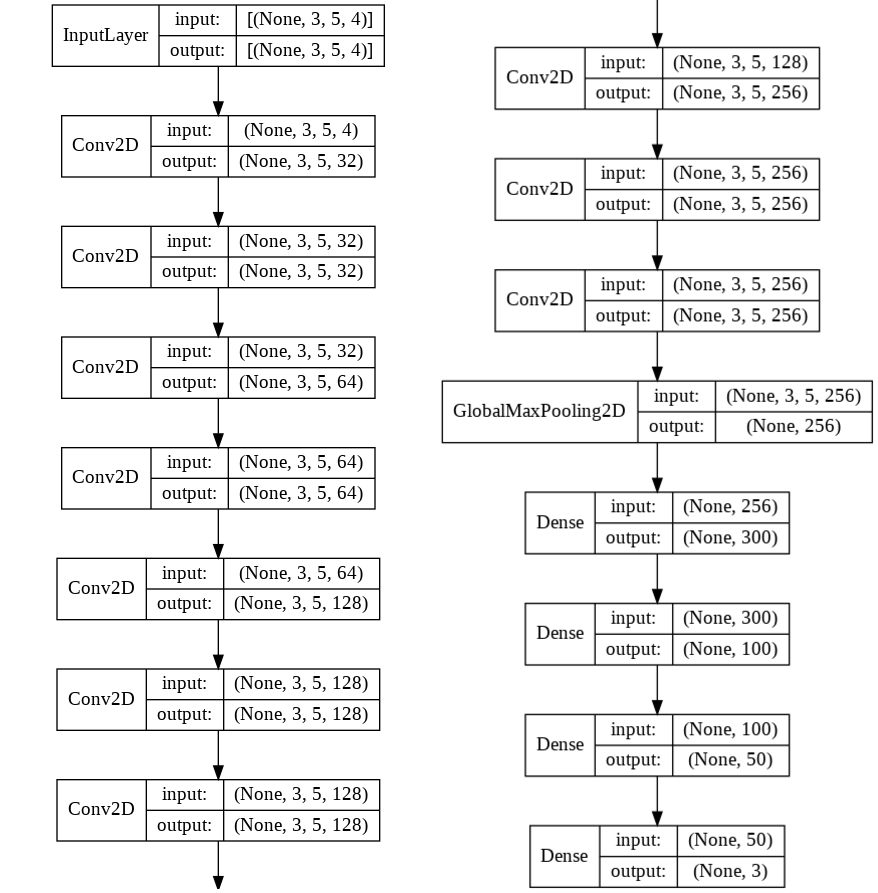}
	\caption{Representation of the CNN for the illustrative case of $N=3$ qubits and circuit depth $P=5$. Boxes report the input and the output shapes of each  layer. This example includes $N_c=10$ convolutional layers (Conv2D). Their  input and output shapes have dimensions: ($B$, $L_1$, $L_2$, $F$), where $B=\mathrm{None}$ is the mini-batch size (not specified), $L_1=N$ and $L_2=P$ denote the size of the two-dimensional feature maps, while $F$ is the number of filters. For the dense layers (Dense),  we only have  the mini-batch size and the number of nodes. The figure omits the batch normalization layers, included in every layer before the application of the activation function. The latter corresponds to the \textit{Mish} function~\cite{Mish}, except for the last node where it corresponds to the \textit{sigmoid} function. It is worth highlighting the global maximum pooling layer (GlobalMaxPooling2D) connecting the last convolutional layer to the first dense layer.}
	\label{summary1}
\end{figure}

The CNNs considered in this article have the overall architecture described in Fig.~\ref{summary1}.
Relevant variations occur  in the last layer. Therein, the number of neurons corresponds the desired number of outputs $N_o$.
Specifically, we consider CNNs designed to predict only one expectation value at a time ($N_o=1$), as well as $N$ expectation values simultaneously ($N_o=N$).
Clearly, in the first case, the last layer includes only one neuron. In the second case, it includes $N$ neurons.
The overall network structure we adopt is standard in fields such as, e.g., image classification or object detection. 
The first part includes $N_c$ multi-channel convolutional layers, which create filtered maps of their input. 
As in standard CNNs, the output of the convolutional part is connected to a few dense layers (four in our case) with all-to-all interlayer connectivity.
However, in standard CNNs, the connection is commonly performed through a flatten layer.
This choice would force to scale the width of the map passed to the first dense layer with the size of the network input.
Therefore, the whole network would be applicable only to one circuit size.
Instead, we perform the connection using a global pooling layer. This extracts the maximum values of each (whole) map in the last convolutional layer.
This feature was adopted in Ref.~\cite{ML_Q5} for the supervised learning of ground-state energies of quantum systems.
It allowed implementing scalable CNNs, i.e., networks that can be trained on heterogeneous datasets including different system sizes and that can predict properties for sizes larger than those included in the training set.
Here, we exploit the global pooling layer to allow a single CNN addressing different circuit sizes.
Notice, however, that full scalability is obtained only when the CNN predicts only one expectation value (with a single neuron in the output layer) at a time. More expectation values corresponding to different qubits can, in fact, be predicted even by the single output network. However, these predictions have to be performed in a sequential manner, by feeding  the network with an appropriate swapping of the features. 
Specifically, when the goal is to predict, say, $z_j$, for any $j\in[2,N]$, one can employ a CNN trained to predict $z_1$, swapping the rows $1$ and $j$ of the network's input.
If, instead, the goal is to simultaneously predict the expectation values corresponding to all qubits, obviously the number of neurons in the last layer has to be adapted to the targeted qubit number. In this case, full scalability is lost.

The training of the CNN is performed by minimizing the loss function. We adopt the binary cross-entropy:
\begin{eqnarray}\label{Loss}
    \mathcal{L}&=&-\frac{1}{N_{\mathrm{train}}}\sum_{k=1}^{N_{\mathrm{train}}}\sum_{i=1}^{N_o}\left[y_i^{(k)}\log(\hat{y}_i^{(k)})\right. \nonumber \\
     &+& \left. (1-y_i^{(k)})\log(1-\hat{y}_i^{(k)})\right],
\end{eqnarray}
where $N_{\mathrm{train}}$ is the number of instances in the training set, $N_o$ is the number of outputs (corresponding to the number of nodes in the last dense layer), and $\hat{y}_i^{(k)}$ is the network prediction corresponding to the (ground-truth) target value $y_i^{(k)}$. 
As discussed above, we consider the cases $N_o=N$ and $N_o=1$. In the first case, the $N$ target values correspond to all rescaled single-qubit expectation values: $y_i=z_i$, for $i=1,\dots,N$.
In the latter case, we consider only one (rescaled) single-qubit expectation value, namely, $y_1=z_1$, or one (rescaled) two-qubit expectation value, namely, $y_1=z_{12}$. 
The optimization method we adopt is a successful variant of the stochastic gradient-descent algorithm, named Adam~\cite{kingma2014adam_OLD}.
No benefit is found by introducing a regularization term in the loss function.
Instead, batch normalization layers are included after every layer, before the application of the activation function. 
The chosen mini-batch size is in the range $N_b\in [128,512]$, depending on the training set size. 
%
The CNNs are trained on large ensembles of random quantum circuits. These are implemented and simulated using the Qiskit library. These simulations provide numerically exact predictions for the considered expectation values.
We also consider noisy estimates of the expectation values obtained from finite samples of simulated measurements. These  estimates are employed in subsection~\ref{subsectionnoise} to inspect the impact of errors in the training set on the prediction accuracy. 
After training, the CNNs are tested on previously unseen random circuits. It is also worth mentioning that we remove possible circuit replicas, both in the training and in the test sets. In fact, only for the smallest circuits we consider, corresponding to $N=3$ and $P=5$, one can find within $N_{\mathrm{train}}\simeq 10^6$ training circuits a non-negligible number of identical replicas.




%
%
\begin{figure}[]
	\centering
	\includegraphics[width=\columnwidth]{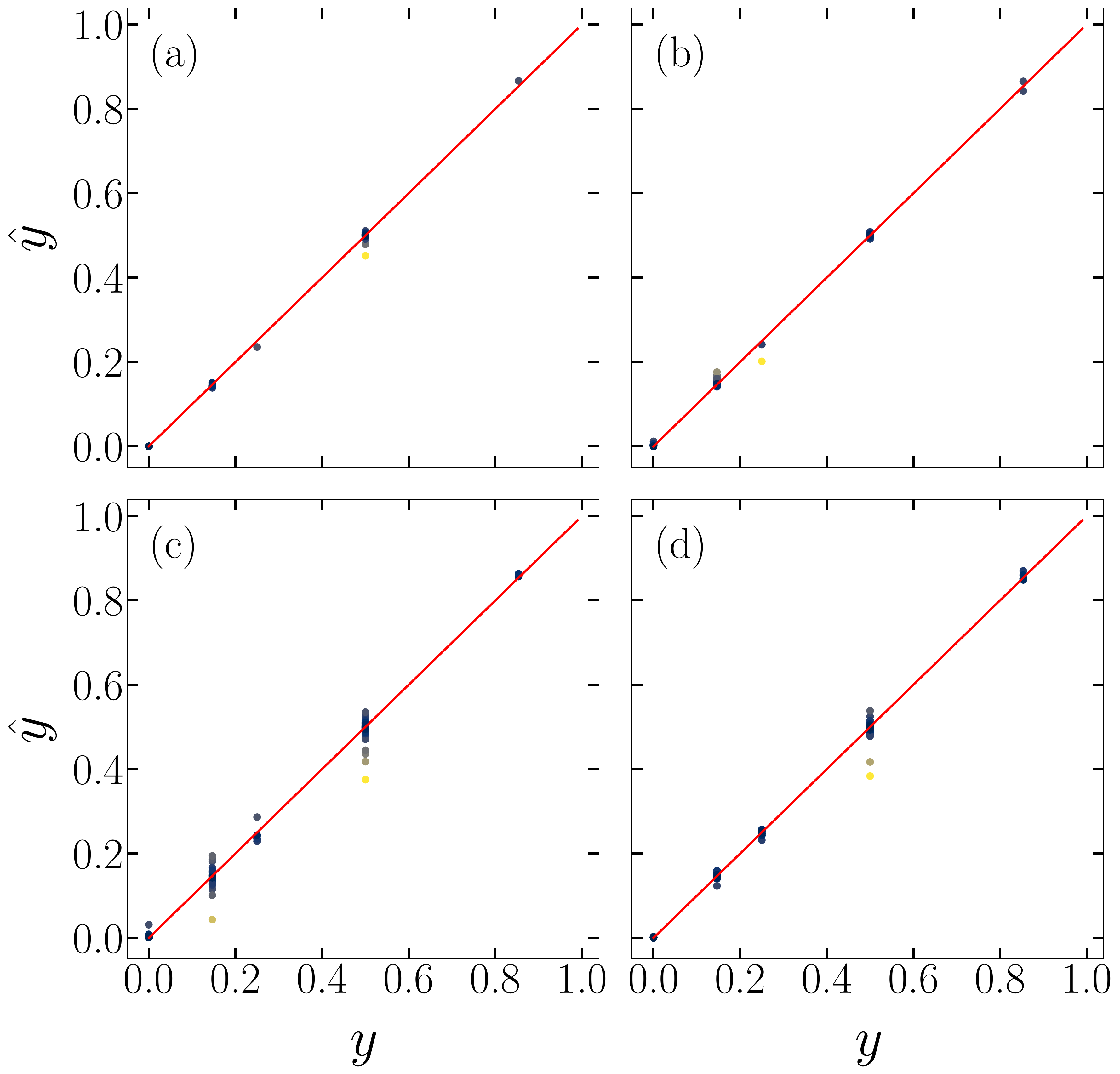}
	\caption{Rescaled single-qubit expectation values $\hat{y}$ predicted by the CNN versus the ground-truth results $y\equiv z_i$, with $i=1,\dots,N$. The latter results are simulated via Qiskit.
	Different number of qubits $N$ and circuit depths $P$ are considered in the four panels: $N=3$ and $P=5$ (a); $N=3$ and $P=7$ (b); $N=5$ and $P=5$ (c); $N=5$ and $P=7$ (d).
	These test sets include 100 random circuits. The color scale (blue to yellow)  represents the absolute discrepancy $\left|\hat{y}-y\right|$. The (red) line represents the bisector $\hat{y}=y$.
	}
	\label{scatter}
\end{figure}

\section{Results}\label{Sec4}
\subsection{Single-qubit expectation values}
\label{singlequbit}
\begin{figure}[!h]
	\centering
	\includegraphics[width=\columnwidth]{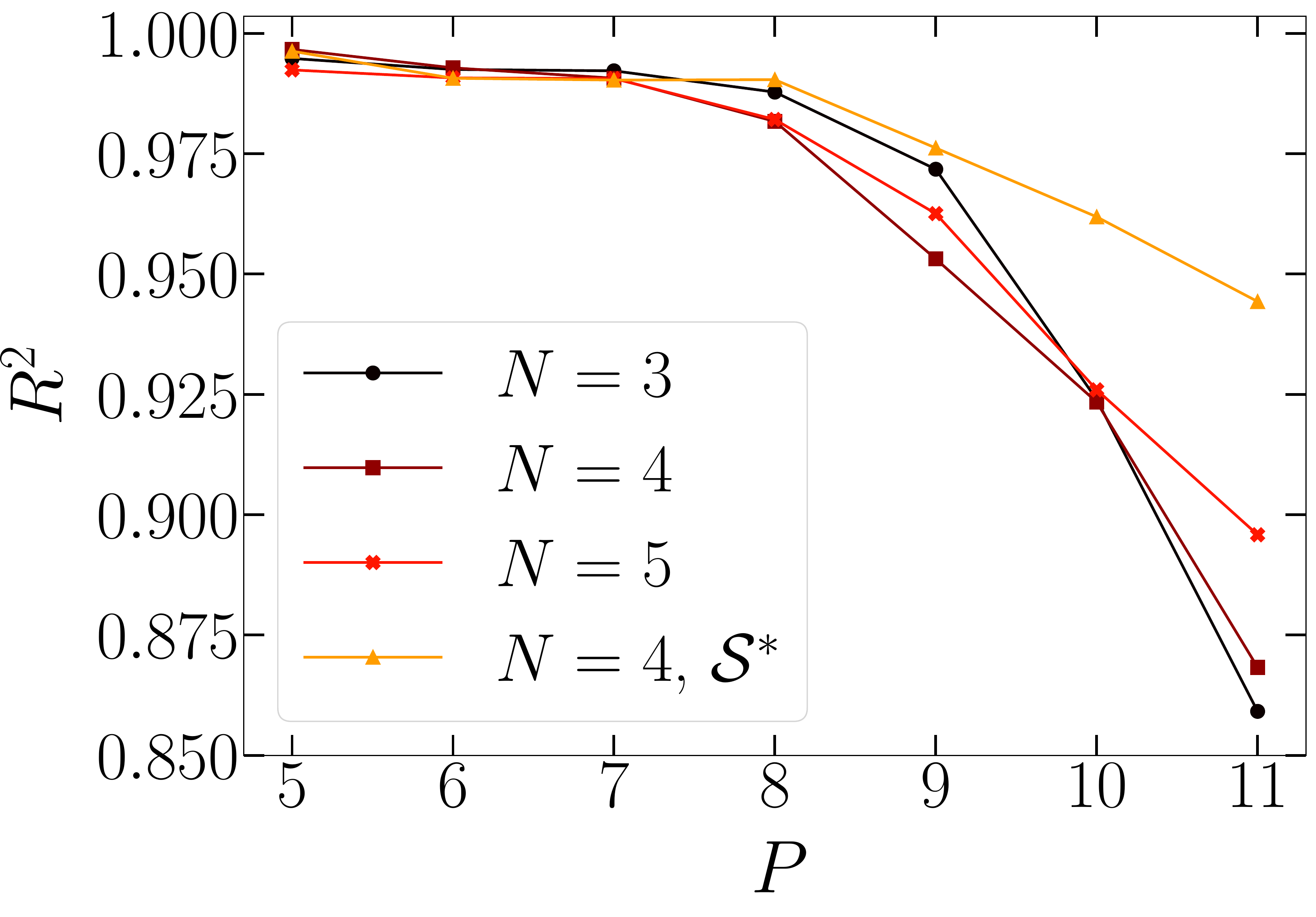}
	\caption{Coefficient of determination $R^2$ for rescaled single-qubit expectation values $z_i$ as a function of the circuits depth $P$.
	Three datasets correspond to random circuits built with gates from the set $\mathcal{S}$, but having different qubit numbers $N$ (see legend). For the fourth dataset, the extended gate set $\mathcal{S}^*$ is used.
	For $P=5$ the neural network is trained from scratch on $N_{\mathrm{train}}\simeq 10^6$ instances, while for $P>5$ the training starts with the optimized weights and biases for $P-1$.
	}
	\label{AN345_z1}
\end{figure}

\begin{figure}[!h]
	\centering
	\includegraphics[width=\columnwidth]{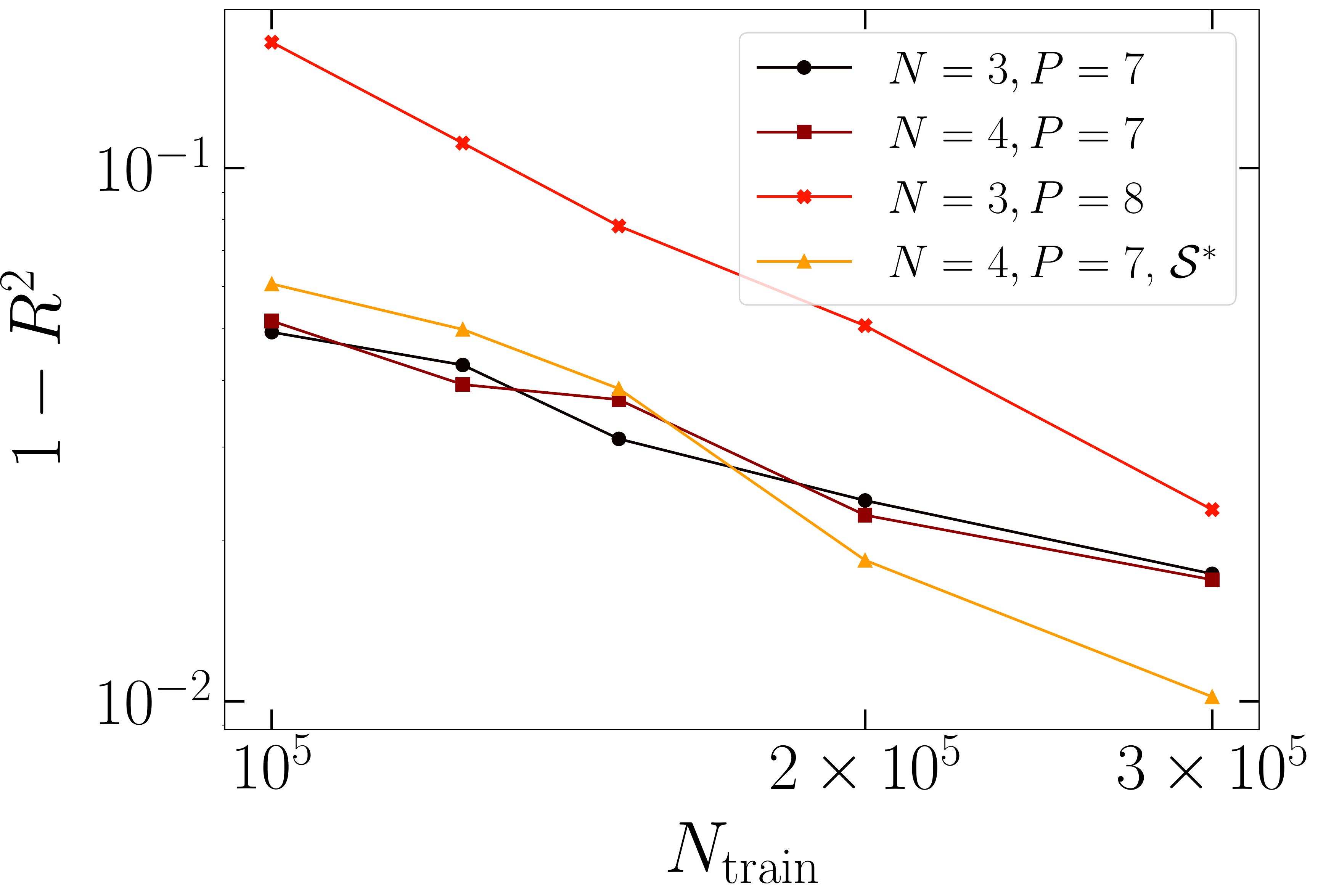}
	\caption{Normalized prediction error $1-R^2$ for (rescaled) single-qubit expectation values $z_i$ as a function of the training set size $N_\mathrm{train}$.  Three datasets correspond to random circuits built with gates from the set $\mathcal{S}$, but having different qubit numbers $N$ and circuit depths $P$ (see legend). For the fourth dataset, the extended gate set $\mathcal{S}^*$ is used. The adopted CNN is described in Fig.~\ref{summary1}.
	}
	\label{trainsize}
\end{figure}

\begin{figure}[!h]
	\centering
	\includegraphics[width=\columnwidth]{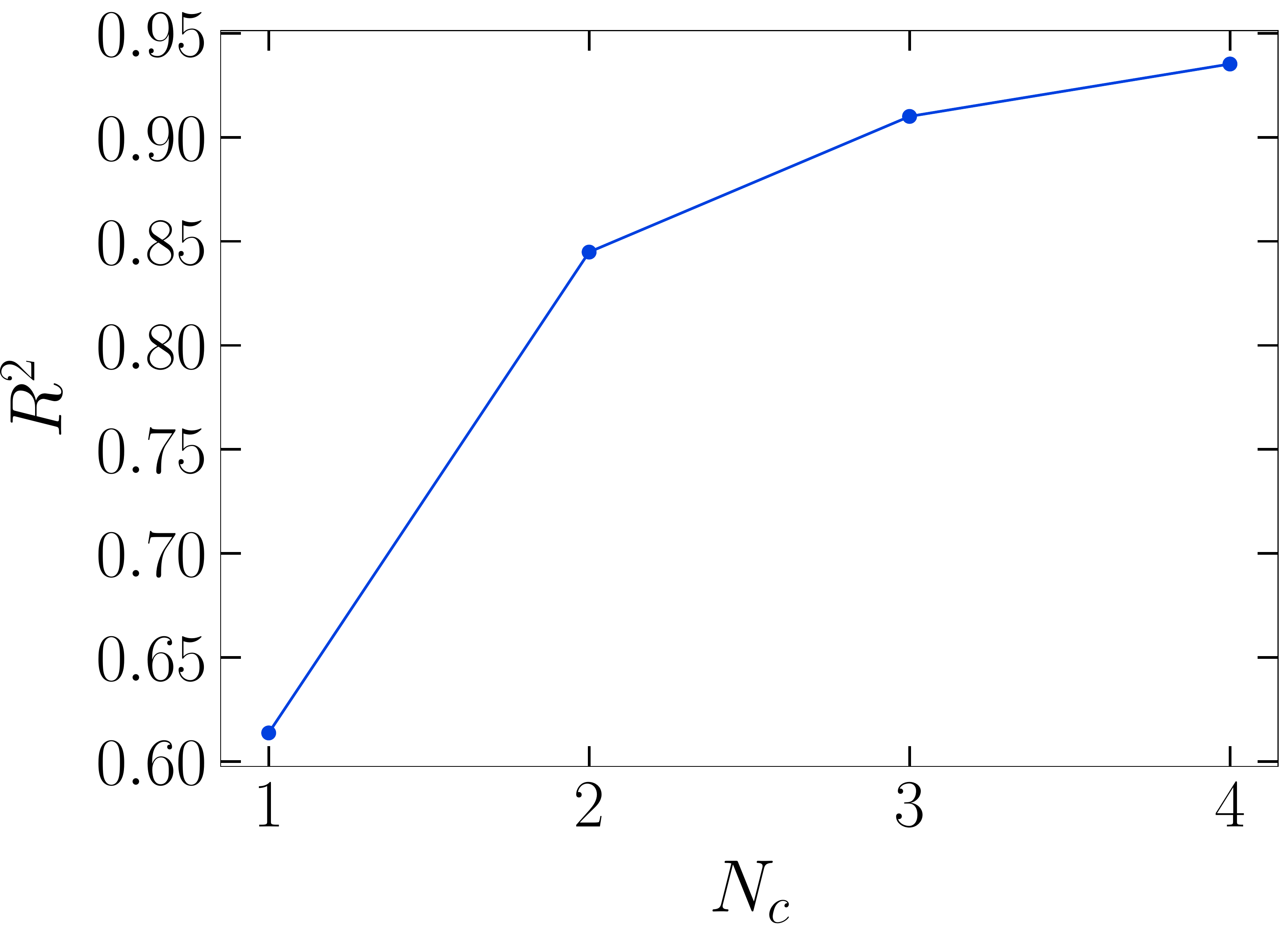}
	\caption{Coefficient of determination $R^2$ for rescaled single-qubit expectation values $z_i$ as a function of the the number of convolutional layers $N_c$. The qubit number is $N=3$  and the circuit depth is $P=7$ .  The training set includes $N_{\mathrm{train}}\simeq 10^6$  random circuits. The employed CNN is similar to the one depicted in Fig.~\ref{summary1}, but with two fully-connected layers including 100 and 50 neurons. The $N_c$ convolutional layers include $F=32$  filters. 
	The batch normalization layers and the output layer act as described in Fig.~\ref{summary1}.
	}
	\label{convs}
\end{figure}

The CNNs described in subsection~\ref{cnnsubsection} are trained to map the circuit descriptors (see subsection~\ref{representationsubsec}) to various outputs.
We first focus on a CNN designed to simultaneously predict the rescaled single-qubit expectation values $z_i$, with $i=1,\dots,N$.
Fig.~\ref{scatter} displays the scatter plots of predicted versus ground-truth expectation values, for four representative circuit sizes. 
The tests are performed on random circuits distinct from those included in the training set.
One notices a close correspondence in all four examples.
In Fig.~\ref{AN345_z1}, we analyse how the prediction accuracy varies with the circuit depth $P$.
Three datasets correspond to random circuits generated with the gates from the set $\mathcal{S}$, but with different qubit numbers $N$. For the fourth dataset the gates are sampled from the extended set $\mathcal{S}^*$ (see subsection~\ref{representationsubsec}).
To quantify the accuracy, we consider the coefficient of determination: 
\begin{equation}\label{r2}
	R^2=1-\frac{\sum_{k=1}^{N_\mathrm{test}}\sum_{i=1}^{N_o}{(y_i^{(k)} - \hat{y}_i^{(k)})^2}}{\sum_{k=1}^{N_\mathrm{test}}\sum_{i=1}^{N_o}{(y_i^{(k)} - \bar{y})^2}}\,
\end{equation}
where $\hat{y}_i^{(k)}$ is the prediction associated to the ground-truth target value $y_i^{(k)}$, $\bar{y}$ is the average of the target values, $N_{\mathrm{test}}$ is the number of random circuits in the test set, and $N_o$ is the number of outputs.
In this case, we have $N_o=N$ outputs.
It is worth stressing that $R^2$ quantifies the accuracy in relation to the intrinsic variance of the test data. It is therefore suitable for fair comparisons among different circuits sizes, which generally correspond to different variances of expectation values.
For small $P$ one observes remarkably high scores $R^2 \simeq 1$, corresponding to essentially exact predictions. However, the accuracy significantly decreases for deeper circuits.
It is worth pointing out that, in this analysis, the depth of the CNN is not varied, and the training set size is also fixed at $N_{\mathrm{train}}\simeq 10^6$. 
The prediction accuracy can be improved by enlarging the training set, or by increasing the depth of the CNN.
The first approach is analyzed in Fig.~\ref{trainsize}, for three representative circuit sizes.
One notices the typical  power-law suppression (see, e.g., \cite{faber2017prediction,hestness2017deep}) of the prediction error $1-R^2\propto N_{\mathrm{train}}^{-\alpha}$, where $\alpha>0$ is a non-universal exponent.
The second approach is analyzed in Fig.~\ref{convs}. We find that the $R^2$ score systematically increases with the number of convolutional layers $N_c$.
It is worth pointing out that the deepest CNNs adopted in this article include around $2\times 10^6$ parameters. 
This number does not represent a noteworthy computational burden for modern high-performance computers, in particular if equipped with state-of-the-art graphic processing units.
In fact, significantly deeper neural networks are routinely trained in the field of computer vision. Relevant examples are VGG-16, VGG-19~\cite{VGG} and ConvNeXt~\cite{convnext}.
Instead, creating copious training sets for circuits with $N>10$ qubits becomes computationally expensive. In fact, simulating circuits with $N\gg10$ qubits is virtually impossible. 
Two strategies to circumvent this problem are discussed in the subsection \ref{subsectiontransfer}.

\subsection{Transfer learning and extrapolation}
\label{subsectiontransfer}
\begin{figure}[!h]
	\centering
	\includegraphics[width=\columnwidth]{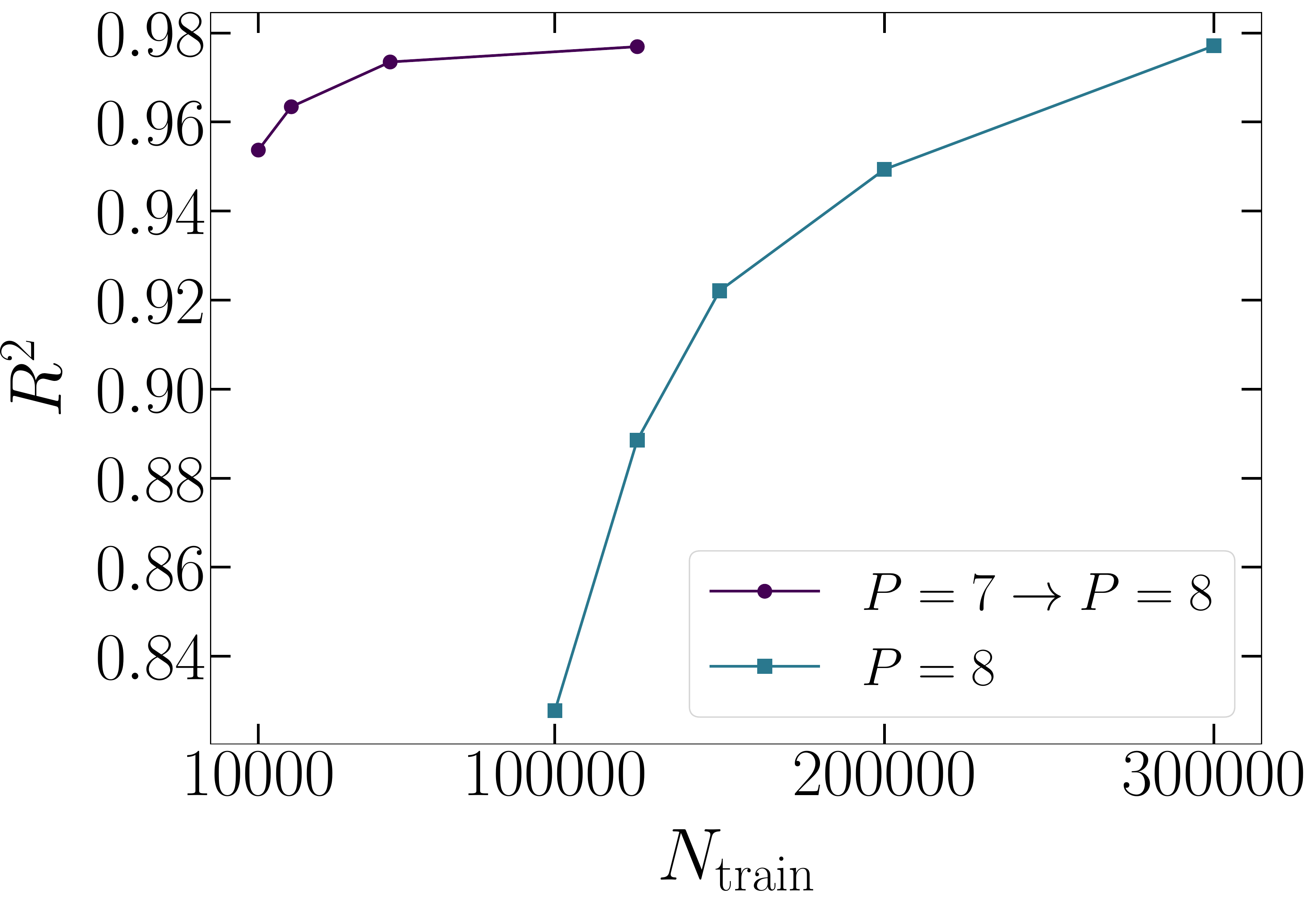}
	\caption{Coefficient of determination $R^2$ for the rescaled single-qubit expectation values $z_1$ as a function of the training set size $N_{\mathrm{train}}$. The size of the test circuits is: $N=3$ and $P=8$. (Blue) squares correspond to training from scratch on the same circuit size. The (violet) circles correspond to transfer learning from $P=7$ to $P=8$ (same qubit number). The pretraining on $P=7$ is performed with $N_{\mathrm{train}}\sim 10^6$ circuits. The CNN is as described in Fig.~\ref{summary1}.
	}
	\label{transfer}
\end{figure}

We exploit the scalability of the CNNs featuring the global pooling layer to exploit two strategies, namely, transfer learning and extrapolation. 
The first strategy is rather common in fields such as, e.g., computer vision~\cite{caruana1997multitask}. 
It involves performing an extensive training of a deep CNN on a large generic database. Then, the pre-trained network is used as starting point in a second training  performed on a more specific, typically smaller, database. At this stage the CNN learns to solve the targeted task.
This approach has already been adopted for the supervised learning of ground-state properties of quantum systems~\cite{10.21468/SciPostPhys.10.3.073,mills2019extensive,ML_Q5,jungsize,scherbela2022solving}.
Here, we use it to accelerate the learning of deep quantum circuits, exploiting a pre-training performed on computationally cheaper circuits with fewer gates per qubit.
Specifically, we compare the learning speed of a CNN trained from scratch on circuits of depth $P=8$, with the one of a CNN pre-trained on circuits of depth $P=7$.
The results are analyzed in Fig.~\ref{transfer}. We find that the pre-trained CNN needs a significantly smaller training set  to reach high $R^2$ scores.

\begin{figure}[!h]
	\centering
	\includegraphics[width=\columnwidth]{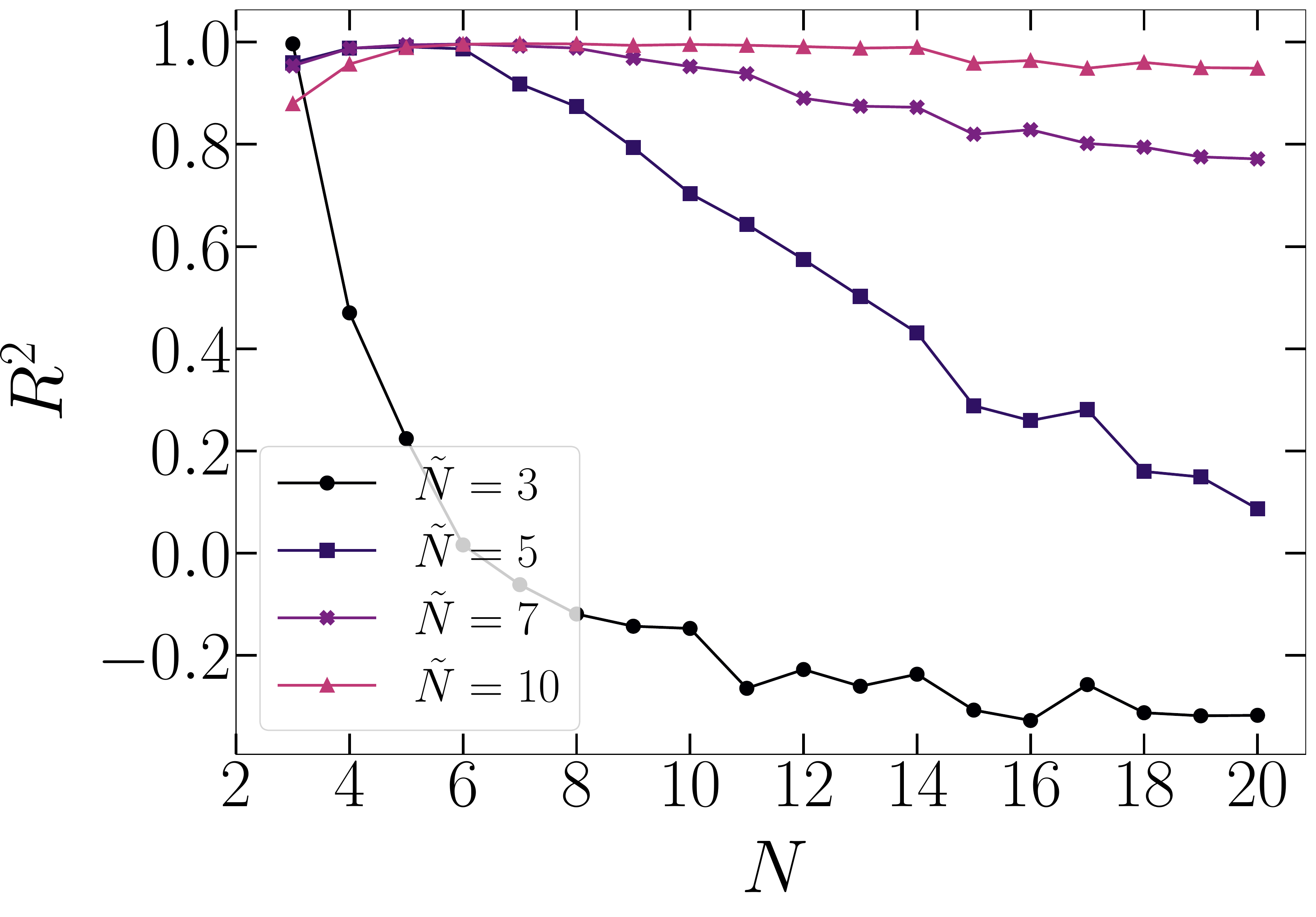}
	\caption{Coefficient of determination $R^2$ for rescaled single-qubit expectation values $z_1$ as a function of the number of qubits $N$ in the test circuits. Both training and test circuits have depth $P=6$. Different datasets correspond to different number of qubits $\tilde{N}$ in the training circuits (see legend). The employed CNN is as shown in Fig.~\ref{summary1} except for the last layer, which has only one neuron. 
	}
	\label{estrap}
\end{figure}

The extrapolation strategy aims at predicting properties of circuits including more qubits than those included in the training set. 
As discussed in subsection~\ref{cnnsubsection}, to allow flexibility in the number of qubits $N$, we adopt the CNN with one output neuron.
This, in combination with the global pooling layer, provides  the network full scalability, allowing the same network parameters to be applied to different circuit sizes.
The results are  analyzed in Fig.~\ref{estrap}. Remarkably, we find that a CNN trained on (computationally affordable) circuits with $\tilde{N}=10$ qubits accurately predicts the (single qubit) expectation values of significantly larger circuits (with $N=20$ qubits). Instead, when the training is performed on smaller circuits ($\tilde{N} \simeq 5$), the $R^2$ score rapidly drops as the test-circuit size increases. This suggests that a minimum training circuit-size is needed to allow the CNN learning how to perform accurate extrapolations.

\subsection{Real quantum computers and noisy simulators}
\label{subsectionnoise}
\begin{figure}[!h]
	\centering
	\includegraphics[width=\columnwidth]{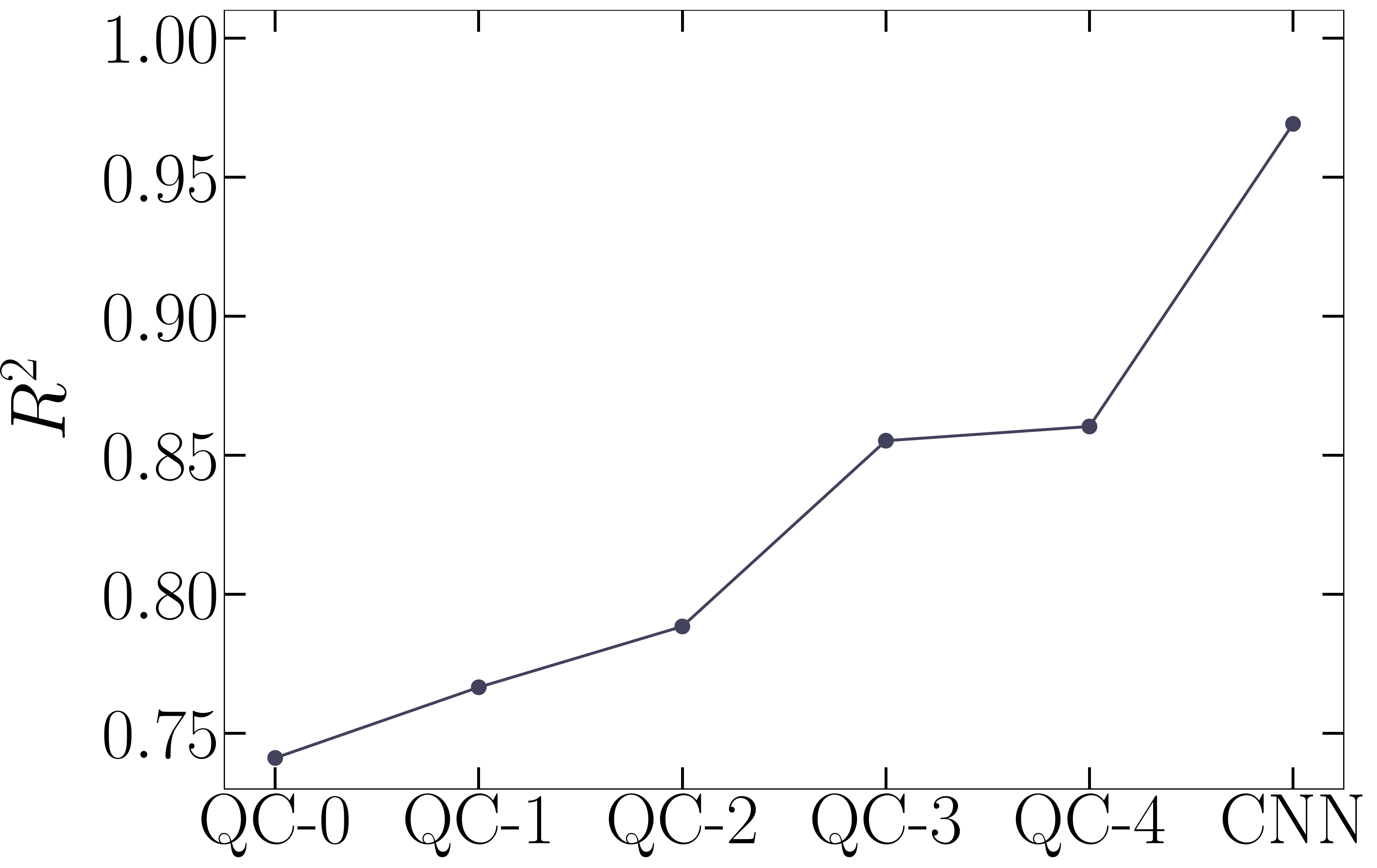}
	\caption{Coefficient of determination $R^2$ for single-qubit expectation values measured on five IBM quantum computers and predicted by a trained CNN. The circuit size is: $N=5$ and $P=10$. The five quantum computers, namely, $\mathrm{QC}\text{-0}\equiv \mathrm{ibmq\_lima}$, $\mathrm{QC}\text{-1}\equiv \mathrm{ibmq\_bogota}$, $\mathrm{QC}\text{-2}\equiv  \mathrm{ibmq\_belem}$, $\mathrm{QC}\text{-3}\equiv\mathrm{ibmq\_quito}$, and $\mathrm{QC}\text{-4}\equiv\mathrm{ibmq\_manila}$, are ordered for increasing $R^2$ score. For each quantum circuit, the expectation values are estimated using $N_{\mathrm{measure}}=2048$ measurements. The training set size is $N_\mathrm{train}\simeq10^7$.
	The CNN is as described in Fig.~\ref{summary1}.
	}
	\label{real}
\end{figure}

\begin{figure}[!h]
	\centering
	\includegraphics[width=\columnwidth]{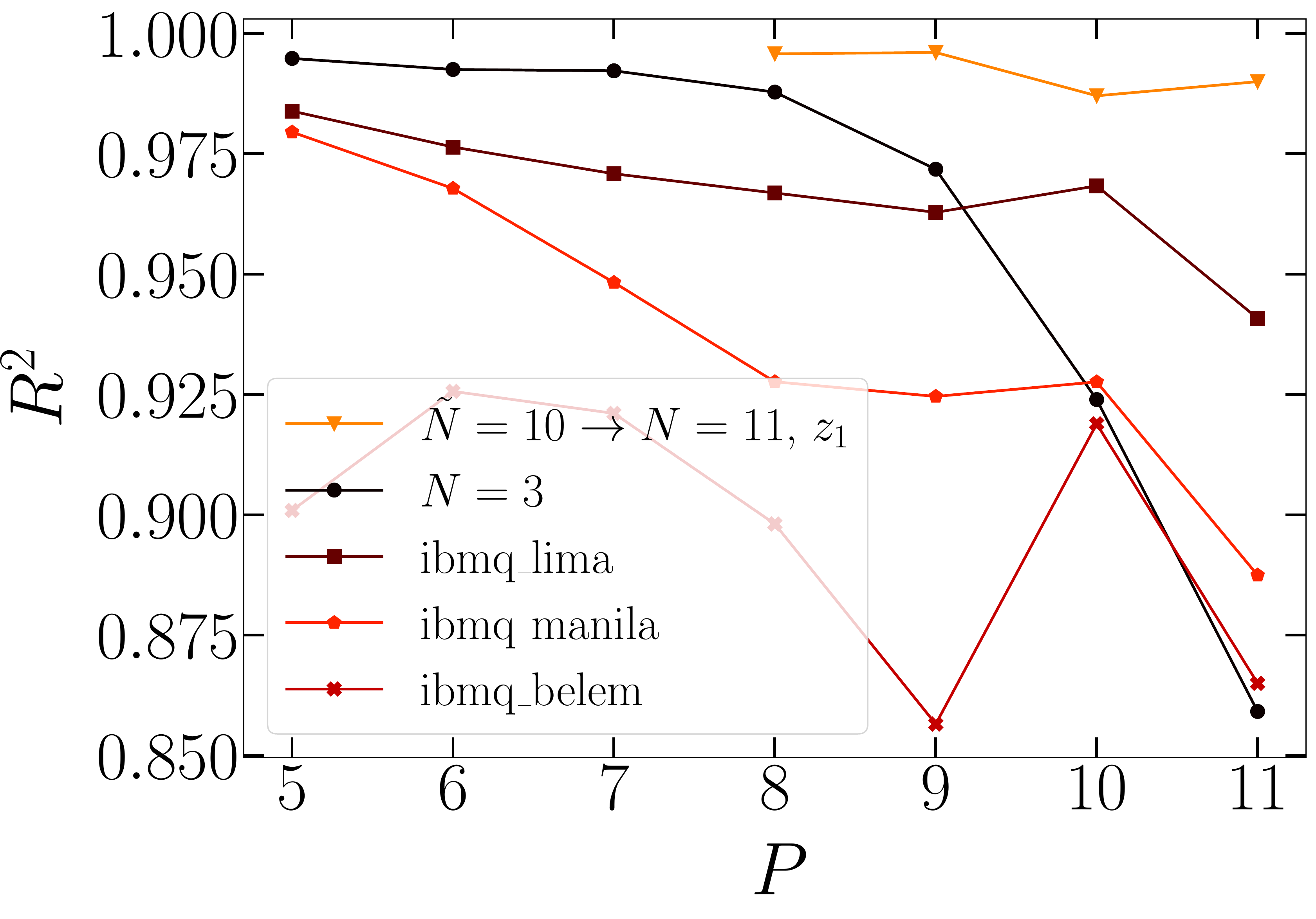}
	\caption{Coefficient of determination $R^2$ for (rescaled) single-qubit expectation values  as a function of the circuit depth $P$. 
	The five datasets correspond to three IBM quantum computers with $N=3$ qubits, and to two CNNs, with $N=3$ and with $N=11$ qubits, respectively.
	The quantum computers are tested on 100 test circuits, estimating the three expectation values via $N_{\mathrm{measure}}=2048$ measurements.
	The CNN for $N=3$ is trained on $N_{\mathrm{train}}\simeq 10^6$ random circuits with the same number of qubits, and it simultaneously predicts the three (rescaled) single-qubit expectation values $z_1$, $z_2$, and $z_3$.
	For $N=11$, the CNN has one output neuron. It is trained on $N_{\mathrm{train}}\simeq 10^7$ circuits with $\tilde{N}=10$ qubits, and it extrapolates to $N=11$ performing a single prediction for $z_1$.
	The training of both CNNs starts with the optimized weights and biases for $P-1$, except for the smallest $P$, where the training starts from scratch.
	}
	\label{real_P}
\end{figure}

Supervised learning with scalable CNNs is being discussed as a potentially useful benchmark for quantum computers. Therefore, it is interesting to compare the predictions provided by trained CNNs with those of actual physical devices.
For this purpose, we execute random circuits on five devices freely available through IBM Quantum Experience~\cite{IBMQ}. 
In Fig.~\ref{real}, the prediction accuracy of a CNN trained on $N_\mathrm{train}\simeq10^7$  classically simulated circuits is compared with the corresponding scores reached by the IBM devices.  The quantum circuits include $N=5$ qubits and $P=10$ gates per qubit. In this case, the neural network outperforms the chosen physical quantum devices.
Another comparison between CNNs and quantum computers is shown in Fig.~\ref{real_P}. Here, only three IBM devices are considered, and the accuracy scores $R^2$ are plotted as a function of the circuit depth $P$, for a fixed number of qubits $N=3$. Notably, for $P>9$, two out of the three  quantum computers outperform the CNN. Notice that the latter is trained on a (fixed) training set with $N_{\mathrm{train}}\simeq 10^6$ instances. 
In fact, it is quite feasible to improve the CNN's accuracy, even for larger qubit numbers. 
As a term of comparison,  we consider in Fig.~\ref{real_P}  also a CNN trained on $N_{\mathrm{train}}\simeq 10^7$ circuits with $N=10$ qubits, and used to extrapolate predictions for $N=11$. One observes that this CNN outperforms all of the considered physical devices. 
We recall that, in Fig.~\ref{estrap} (see also Fig.~\ref{Z1Z2_estrap} below), accurate extrapolations to even more challenging  qubit numbers $N = 20$ are demonstrated.
These findings indicate that scalable CNNs trained via supervised learning on classically simulated quantum circuits represent a potentially useful benchmark for the development of quantum devices.  

\begin{figure}[!h]
	\centering
	\includegraphics[width=\columnwidth]{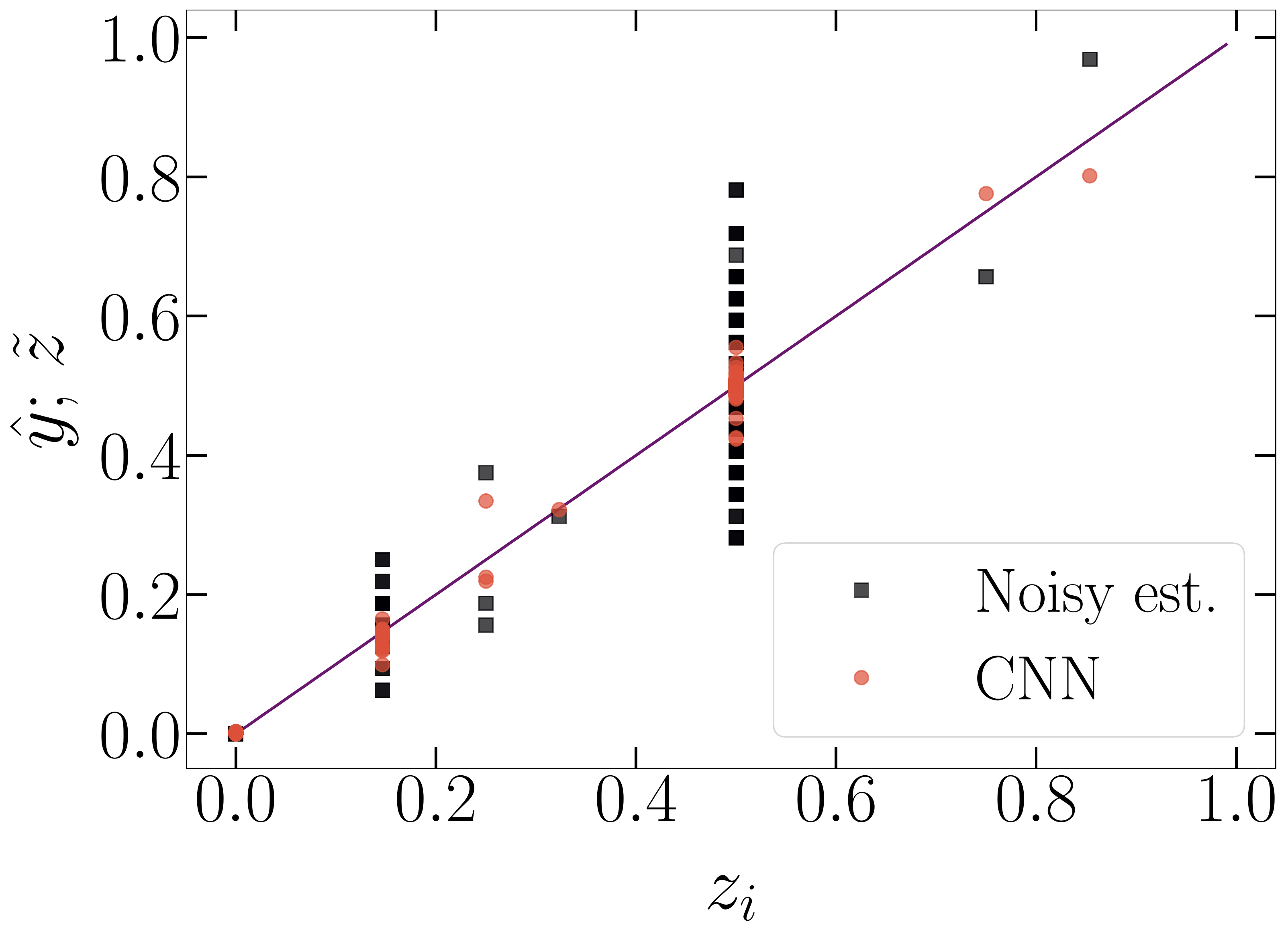}
	\caption{
	Predictions of (rescaled) single-qubit expectation values versus ground-truth results $y\equiv z_i$ (see Eq.~\eqref{totargetvalue}).
	The CNN predictions $\hat{y}$ (red circles) are compared against the noisy estimates $\tilde{z}_i$ (black squares) obtained by averaging $N_{\mathrm{measure}}=32$ (simulated) measurements.
	The circuits size is $N=3$ and $P=7$. The CNN is trained on the noisy estimates corresponding to $N_{\mathrm{train}}\simeq 10^6$ random circuits.
	}
	\label{scatternoise}
\end{figure}

\begin{figure}[!h]
	\centering
	\includegraphics[width=\columnwidth]{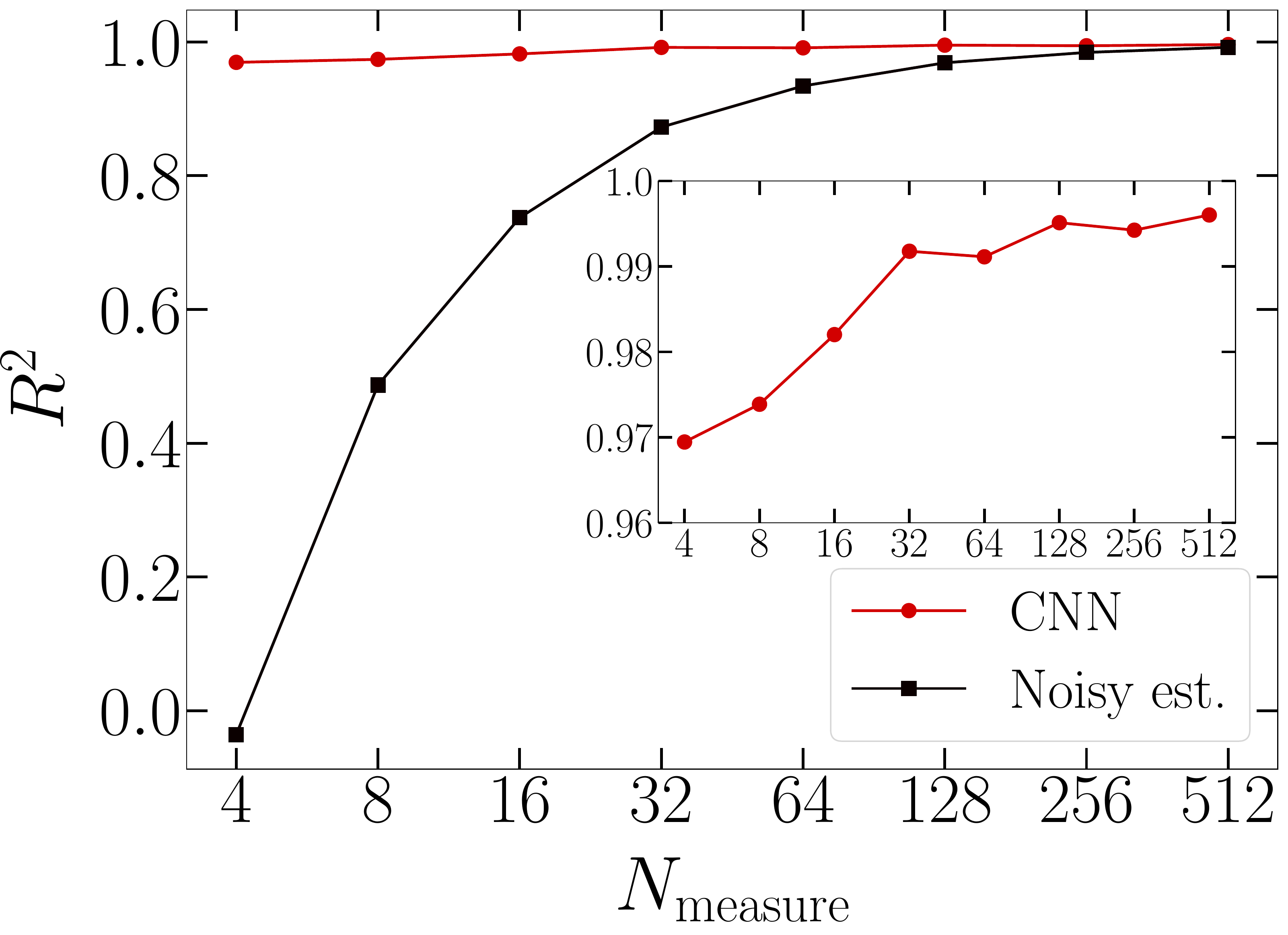}
	\caption{
	Coefficient of determination $R^2$ as a function of the number of simulated measurements $N_{\mathrm{measure}}$. In the main panel, the $R^2$ score of the noisy estimates $\tilde{z}_i$ with respect to the ground-truth (rescaled) single-qubit expectation values $z_i$ (black squares) is compared with the corresponding score of the CNN predictions (red circles). The circuits size is $N=3$ and $P=7$. The inset displays the CNN data on a narrower scale.
	}
	\label{noise}
\end{figure}

One can envision the use of data produced by physical quantum devices to train scalable CNNs. This could allow them learning how to emulate classically intractable quantum circuits. 
However, physical devices only allow estimating output expectation values via finite number of measurements. In the era of noisy intermediate quantum computers~\cite{Preskill2018quantumcomputingin}, one should expect this number to be quite limited, leading to noisy estimates affected by significant statistical fluctuations. 
Therefore, it is important to analyse the impact of this noise on the supervised training of CNNs.
For this, we consider as training target values the noisy estimates obtained by simulating via Qiskit finite numbers of measurements $N_{\mathrm{measure}}$. 
In the testing phase, the CNN's predictions are compared against exact expectation values.
This comparison is shown in the scatter plot of Fig.~\ref{scatternoise}, for the case of $N_{\mathrm{measure}}=32$.
One notices that, while the noisy estimates display large random fluctuations, the CNN's predictions accurately approximate the exact expectation value.
This effect is quantitatively analyzed via the $R^2$ score in Fig.~\ref{noise}.
Notably, the CNN reaches remarkably accuracies $R^2 \gtrsim 0.99$ for numbers of measurements as small as $N_{\mathrm{measure}}\sim 32$, despite the fact that the estimated expectation values are significantly inaccurate, corresponding to $R^2\simeq 0.9$.
An analogous resilience to noise in training data was first observed in applications of CNNs to image classification tasks~\cite{rolnick2017deep}. It was also demonstrated in the supervised learning of ground-state energies of disordered atomic quantum gases~\cite{ML_Q4}.
It is quite relevant to recover this property in the case of quantum computing, where noise represents a major obstacle to be overcome.
Chiefly, this resilience paves the way to the use of quantum computing devices for the production of training datasets. Deep CNNs could be trained to solve classically intractable circuits, and then  distributed to practitioners more easily than a physical device.

\subsection{Emulation of the Bernstein-Vazirani algorithm}
\begin{figure}[!h]
	\centering
	\captionsetup[subfigure]{justification=centering}
	\begin{subfigure}[b]{0.45\columnwidth}
		\includegraphics[width=\columnwidth]{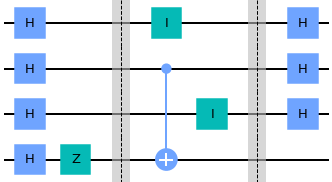}
		\caption{}
	\end{subfigure}
	\hspace{\fill}
	\begin{subfigure}[b]{0.45\columnwidth}
		\includegraphics[width=\columnwidth]{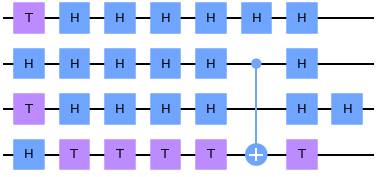}
		\caption{}\label{BVdiagram00}
	\end{subfigure}

	\caption{Representation of the Bernstein-Vazirani (BV) algorithm for a  sought-for string $\textbf{w}\in\{0,1\}^3$, corresponding to a circuit with $N=4$ qubits. Panel (a) displays the conventional implementation using the gates $I$, $Z$, $H$, and $CX$, for the bit string $\textbf{w}=010$. Panel (b) displays an alternative implementation using only gates from the set $\mathcal{S}$. The dangling $T$-gate on the last qubit is required for shape consistency, and it does not affect the relevant output probabilities.
	}
	\label{BVdiagram}
\end{figure}

\begin{figure}[!h]
	\centering
	\captionsetup[subfigure]{justification=centering}
	\begin{subfigure}[]{\columnwidth}
		\includegraphics[width=\columnwidth]{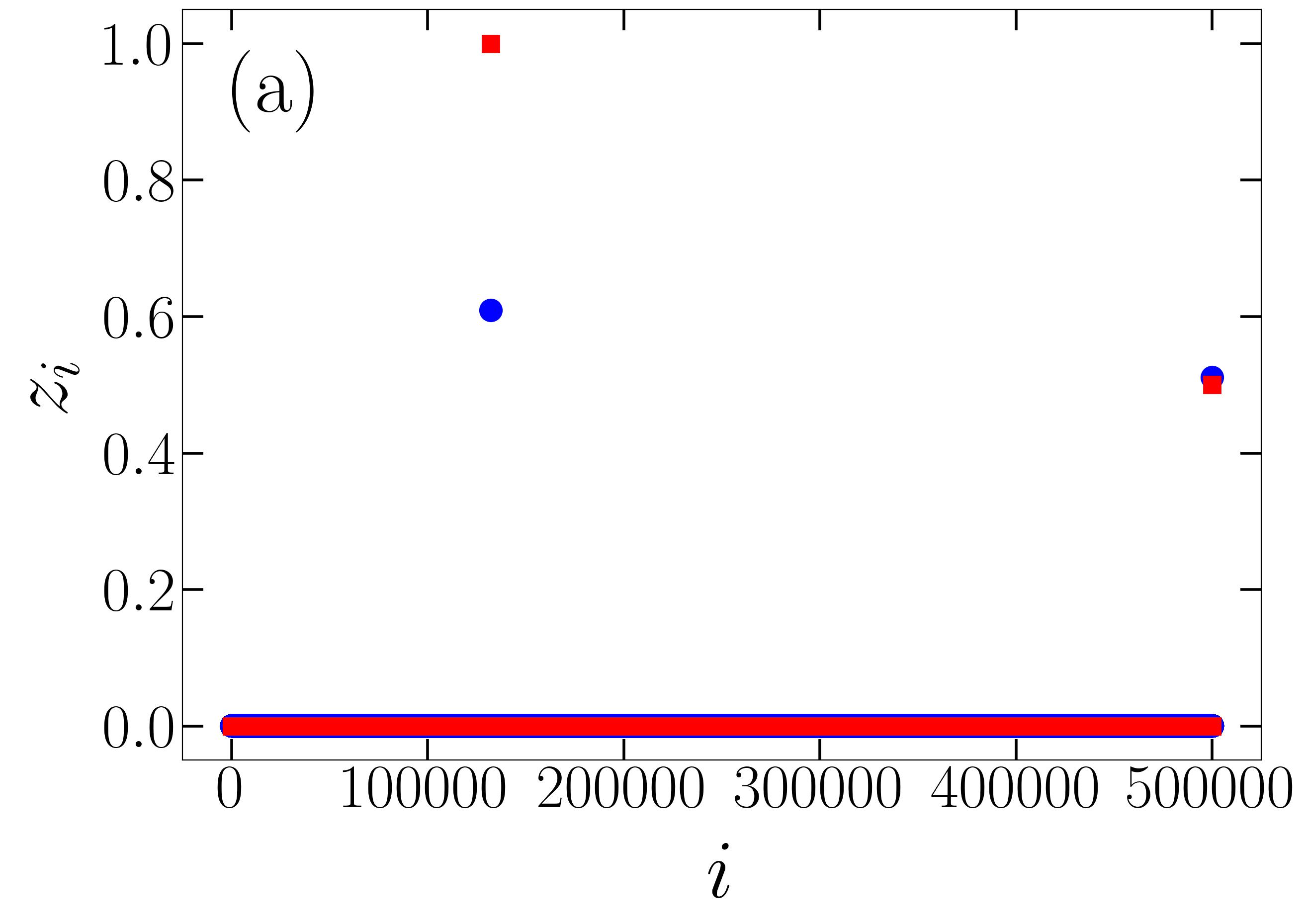}
	\end{subfigure}
	\begin{subfigure}[]{\columnwidth}
		\includegraphics[width=\columnwidth]{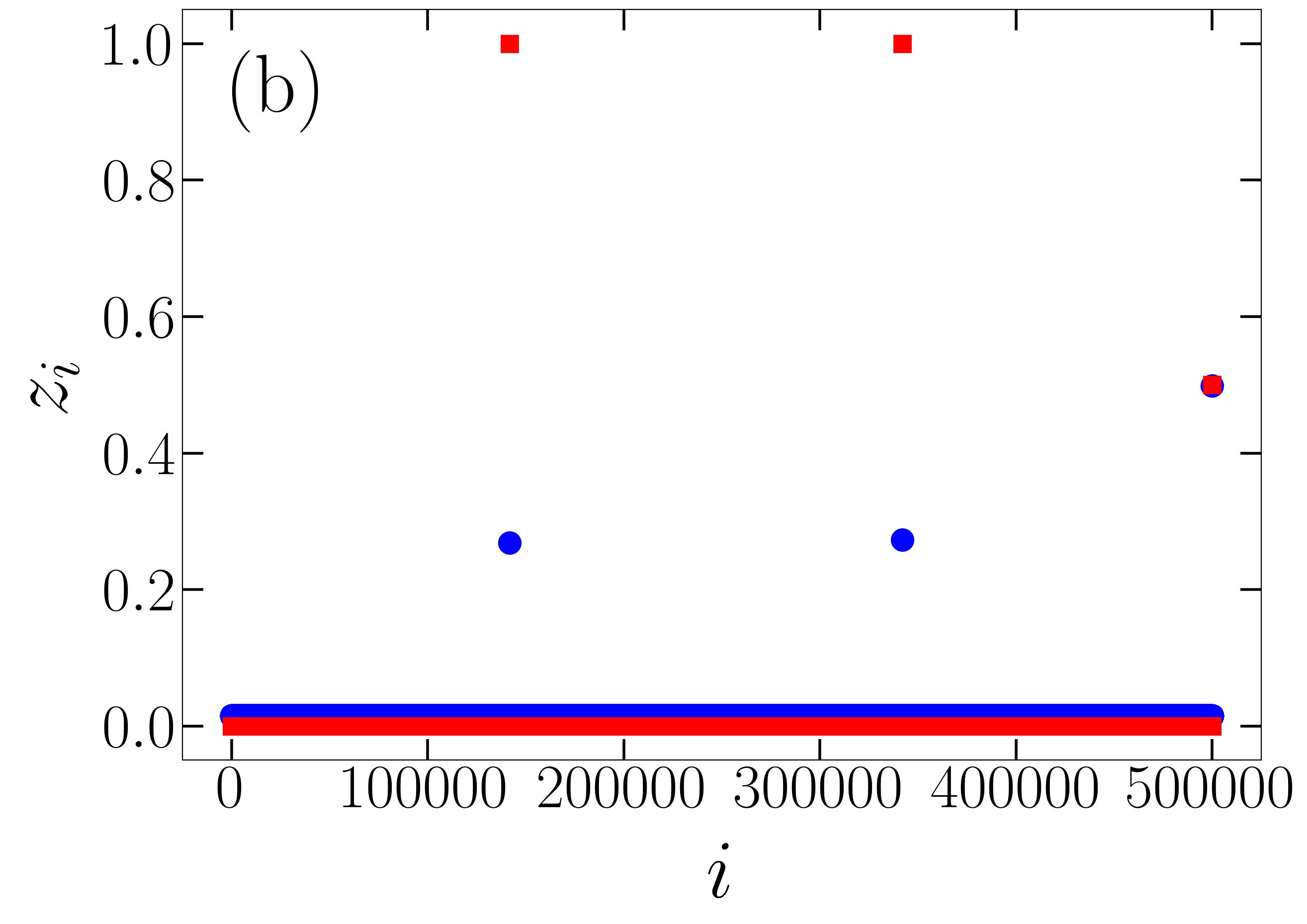}
	\end{subfigure}
	\begin{subfigure}[]{\columnwidth}
		\includegraphics[width=\columnwidth]{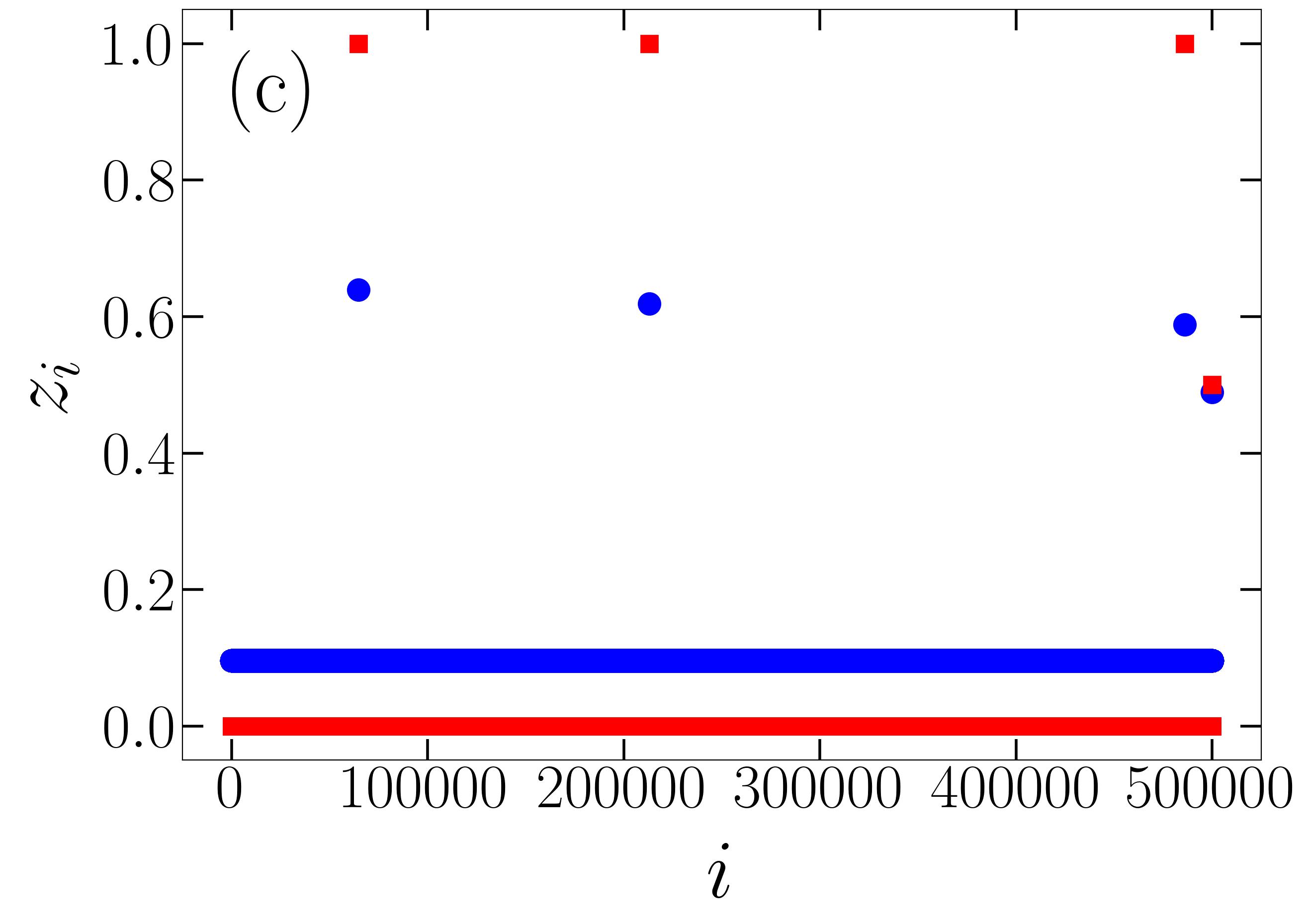}
	\end{subfigure}

	\caption{Rescaled output expectation values $z_i$ (see Eq.~\eqref{totargetvalue}) as a function of the qubit index $i=1,\dots,N$, for a BV algorithm with $N=5\times 10^5$ qubits. 
	The CNN predictions (blue circles) are compared to the expected values (red squares).
	The three panels correspond to sought-for bit strings $\textbf{w}$ with one non-zero bit $i_1=132121$ [panel (a)], with two non-zero bits $(i_1,i_2)=(341924, 141725)$ [panel (b)], and with three non-zero bits $(i_1,i_2,i_3)=(64793, 212973, 485883)$ [panel (c)].
	The CNNs are trained on $N_{\mathrm{train}}\simeq 10 ^6$ [panel (a)] and $N_{\mathrm{train}}\simeq 10 ^7$ [panels (b) and (c)] random circuits with $N=10$ qubits.
	}
	\label{BV_results}
\end{figure}

As discussed in subsection~\ref{emulable}, the BV  algorithm can be emulated by predicting the single-qubit expectation values $\braket{Z_i}$, with $i=1,\dots,N-1$. This means that these expectation values allow one unequivocally identifying the sought-for bit string $\textbf{w}\in \{0,1\}^{N-1}$.
Here we analyse how accurately our scalable CNNs emulate this algorithm. Notably, we challenge the CNN in the extrapolation task, i.e., we use it to emulate BV circuits with (many) more qubits than those included in the training set.
Conventionally, the BV algorithm is implemented using the following gates: $I$, $Z$, $H$, and $CX$. However, it can also be realized using only gates from the set $\mathcal{S}$. 
An example of this alternative implementation is visualized in Fig.~\ref{BVdiagram}.
Notice that a dangling $T$ gate, acting on the $N$-th qubit in the last layer, needs to be inserted. However, this does not affect the relevant output expectation values.
The tests we perform are limited to sought-for bit strings $\textbf{w}$ where all bits except one, two, or three, have zero value; that is, only one, two, or three indices $i_{\alpha}\in[1,N-1]$ exist such that $w_{i_{\alpha}}=1$, where $\alpha\in\{1,2,3\}$ spans the group of (up to three) non-zero bits. These indices are randomly selected.
Specifically, a CNN is trained on circuits with $\tilde{N}=10$ qubits and depth $P=7\text{, }8,\text{ or } 9$, for one, two, or three non-zero bits, respectively.  It is then invoked to predict the single qubit expectation values of BV circuits with larger $N$.
To visualize the prediction accuracy, we show in Fig.~\ref{BV_results} the expectation values $z_i$, for $i\in[1,N]$, for a BV circuit with as many as $N=5\times 10^5$ qubits. 
Notice that also the $N$-th expectation value, corresponding to the ancilla qubit, is shown.
One notices that, beyond the $N$-th qubit, all but one, two, or three expectation values are small, meaning that the CNN is able to identify the sought-for indices $i_{\alpha}$ corresponding to $w_{i_{\alpha}}=1$.
It is remarkable that a CNN trained on random circuits learns to emulate a rather peculiar algorithm such as the BV circuit, even for larger qubit numbers.

\begin{figure}[!h]
	\centering
	\includegraphics[width=\columnwidth]{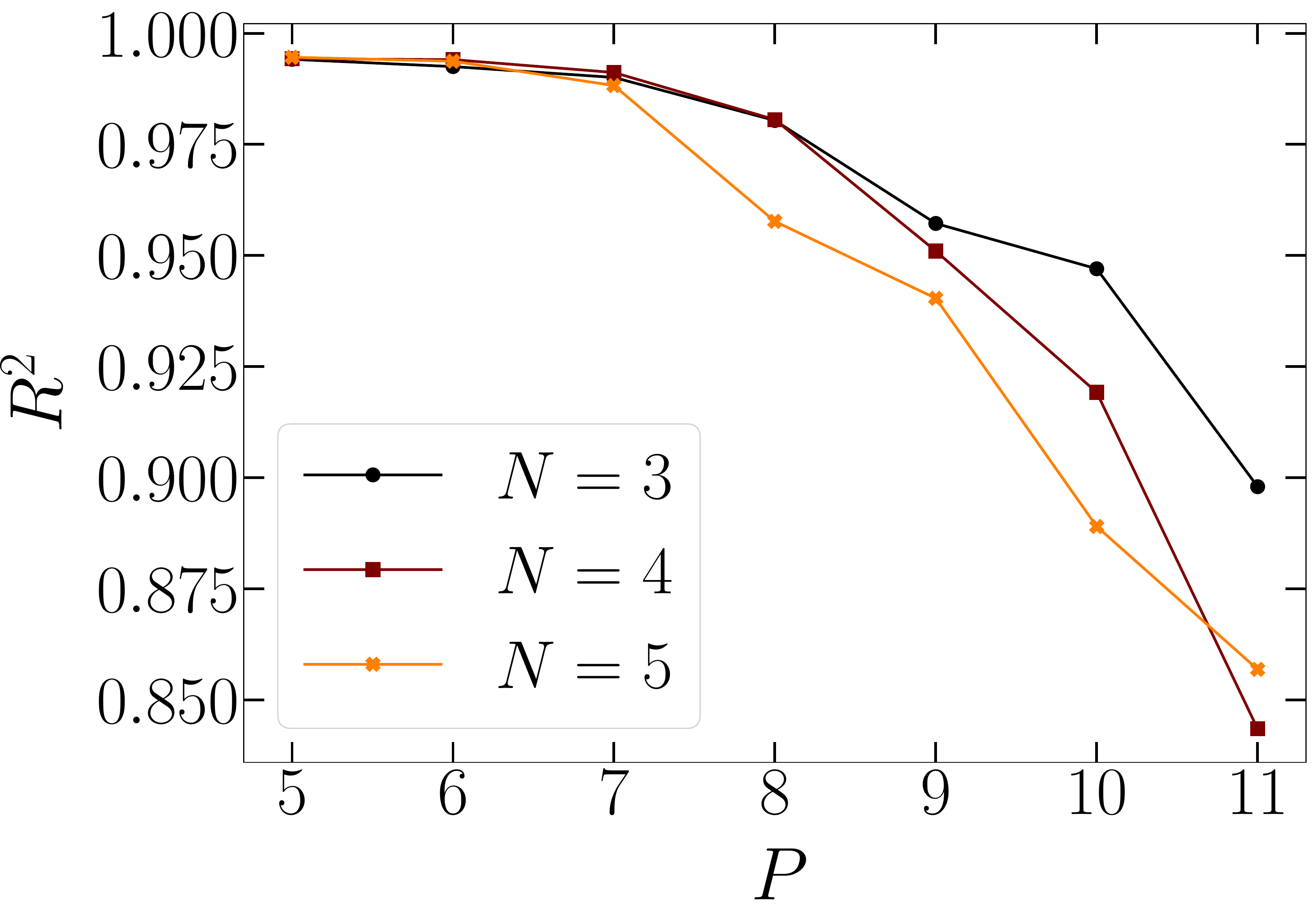}
	\caption{Coefficient of determination $R^2$ for the rescaled two-qubit expectation value $z_{12}$ (see Eq.~\eqref{eqz12}) as a function of the circuit depth $P$. The three datasets correspond to different qubit numbers $N$.
	The training set size is $N_\mathrm{train}\simeq 10^6$. 
	For $P=5$ the CNN is trained from scratch, while for $P>5$ the weights and biases are initialized at the optimized values for $P-1$. This allows a significant reduction in computation time.
	}
	\label{Z1Z2_traintest}
\end{figure}

\subsection{Two-qubit expectation values}
The scalable CNN can be trained to predict also two-qubit expectation values. We consider only the first two qubits, i.e., the CNN predicts the rescaled expectation value $z_{12}$ defined in Eq.~\eqref{eqz12}. 
Henceforth, only one neuron is included in the output layer (see discussion in subsection~\ref{cnnsubsection}).
Again, it is worth pointing out that the same CNN could predict also other two-qubit expectation values, corresponding to any pair $(i,j)$. These predictions are obtained by performing the double exchange of row indices $(1,2) \longleftrightarrow (i,j)$ in the circuit descriptor matrix. 
In Fig.~\ref{Z1Z2_traintest}, the prediction accuracy is analysed as a function of the circuit depth $P$.
%
%
One observes remarkably high scores $R^2\simeq 1$ for small and intermediate circuits depths, and a moderate accuracy degradation for deeper circuits.
As already shown for single-qubit expectation values (see subsection~\ref{singlequbit}), we stress that also in this case the prediction accuracy can be further improved by increasing the training set size or deepening the CNN (data not shown).
%

%

%
\begin{figure}[!h]
	\centering
	\includegraphics[width=\columnwidth]{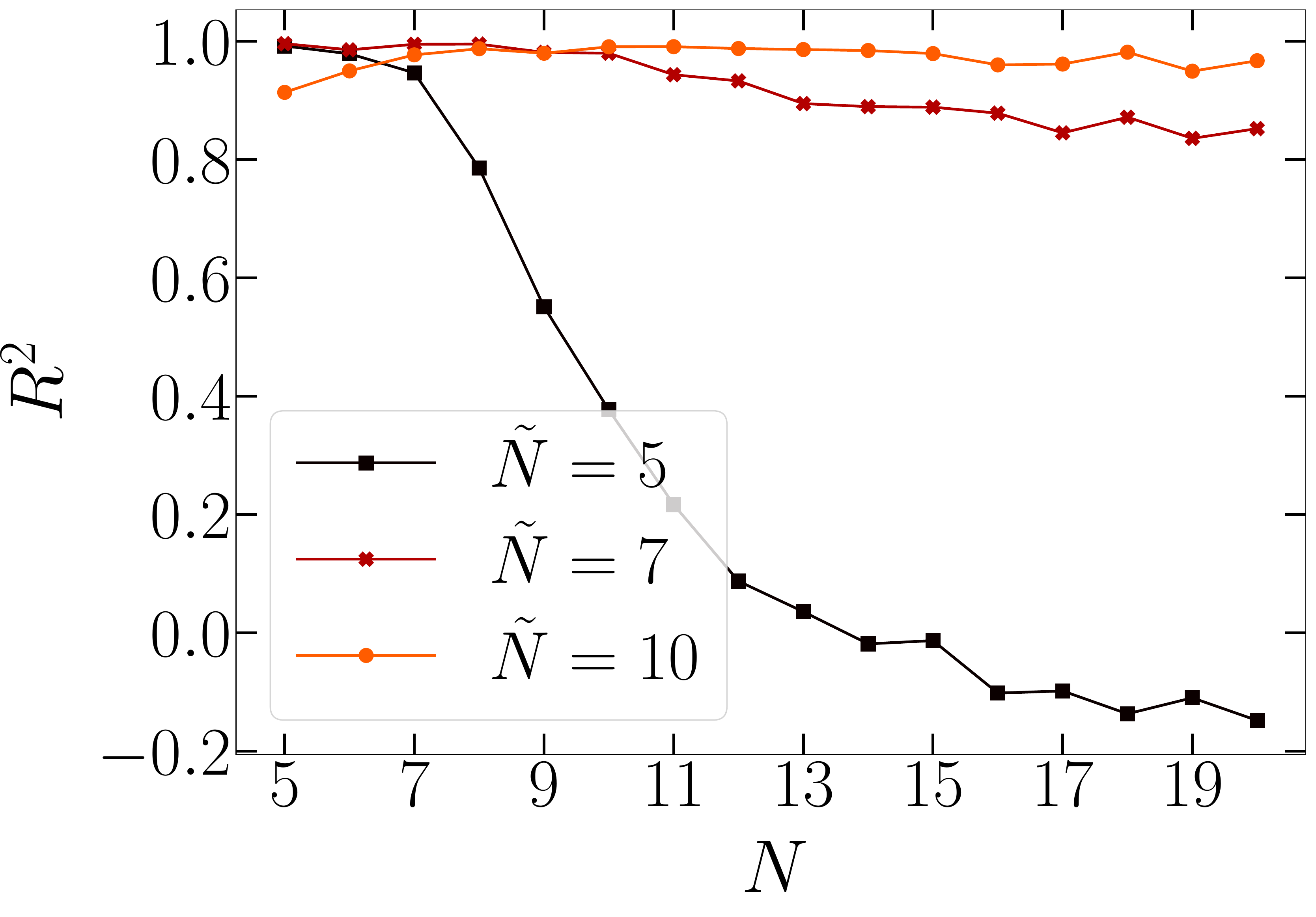}
	\caption{Coefficient of determination $R^2$ for the rescaled two-qubit expectation value $z_{12}$ as a function of the number of qubits $N$ of the test circuits. The three datasets correspond to different qubit numbers $\tilde{N}$ of the training circuits. The training set includes $N_{\mathrm{train}}\simeq 10^6$ random circuits. Both training and test circuits have depth $P=6$.
	}
	\label{Z1Z2_estrap}
\end{figure}

The scalable CNN is also tested in the extrapolation task, i.e., in predicting the two-qubit expectation value for circuits larger than those included in the training set. The accuracy score $R^2$ is plotted in Fig.~\ref{Z1Z2_estrap}. The three datasets corresponds to different training qubit numbers $\tilde{N}=5\text{, }7\text{, and } 10$, and the extrapolation is extended up to $N=20$. 
Notably, if the training qubit number is sufficiently large, the predictions remain remarkably accurate for significantly larger circuits.

\section{Conclusions}\label{Sec5}
We explored the supervised learning of random quantum circuits.
Deep CNNs have been trained to map a properly designed one-hot encoding of the circuit quantum  gates to the corresponding single-qubit and two-qubit output expectation values.
After training on sufficiently large datasets of classically simulated circuits, the CNN provided remarkably accurate predictions, even superior than those provided by small quantum computers available for free from the IBM quantum program.
Notably, we implemented scalable CNNs. This allows them predicting the properties of (computationally challenging) circuit sizes larger than those included in the training set.
This represents a promising strategy to produce benchmark data for classically intractable qubit numbers.
The considered expectation values represent an admittedly limited description of the circuits' output. However, we demonstrated that circuits with only one or two possible output bit strings can be emulated using these targeted expectation values.
Notably, the BV algorithm belongs to this category, and we use it as a relevant benchmark for the CNN's predictions. In fact, we verified that the scalable CNN can emulate it even for qubit numbers much larger than those included in the training set.
The supervised learning turned out to be remarkably robust against random errors in the training set. 
Specifically, the CNN provided accurate predictions for expectation values even when trained on noisy averages performed over few simulated measurements. This finding suggests that CNNs could be trained on data produced by noisy intermediate-scale quantum computers~\cite{Preskill2018quantumcomputingin}. 
To produce the results reported in this article, several training and test datasets have been produced, and various CNNs have been trained. To facilitate future comparative studies, avoiding cluttering the repository, we provide the datasets and the code used for one emblematic test, namely, the one analysed in Fig.~\ref{estrap}, through Ref.~\cite{repository}.

Classical simulations of quantum algorithms play a pivotal role in the development of quantum computing devices.
On the one hand, they provide benchmark data for validation. On the other hand, they represent an indispensable term of comparison to justify claims of quantum speed-up in the solution of computational problems~\cite{troyerdefining}.
For adiabatic quantum computers, quantum Monte Carlo algorithms have emerged as the standard benchmark~\cite{doi:10.1126/science.1068774,boixo2013experimental,boixo2014evidence,troyerheim}. This stems from their ability of simulating the tunneling dynamics of quantum annealers based on sign-problem free Hamiltonians~\cite{PhysRevLett.117.180402,PhysRevB.96.134305,brady2016quantum,PhysRevA.97.032307,PhysRevB.100.214303}.
Simulating universal gate-based quantum computers is more challenging. 
Direct simulation methods, such as those based on tensor networks~\cite{PhysRevLett.126.170603}, are being continuously improved~\cite{jones2019quest,guerreschi2020intel,Steiger2018projectqopensource,villalonga2020establishing,PhysRevLett.128.030501}, but they anyway suffer from an exponentially-scaling computational cost.
Supervised machine-learning algorithms were recently proven to be able of solving  computational tasks which are intractable for algorithms that did not learn from data~\cite{ML_Q3}. In this article, we investigated their efficiency  in simulating a limited description of  quantum circuits' output.
It is worth mentioning that the combination of classical machine learning and quantum computers has already been discussed in various contexts~\cite{Zhang_2021,doi:10.1080/00107514.2014.964942,seif2018machine,PhysRevLett.123.230504}. For example, in Ref.~\cite{PhysRevResearch.2.022060}, generative neural networks trained via unsupervised learning were used to accelerate the convergence of expectation-value estimation. 
One can envision the use of stochastic generative neural network to predict a more complete description of the circuits' outputs such as, e.g., the classical shadow~\cite{ML_Q3, huang2020predicting}. We leave this endeavour to future investigations.

\section*{Acknowledgments} This work was supported by the Italian Ministry of University and Research under the PRIN2017 project CEnTraL 20172H2SC4, and by the European Union Horizon 2020 Programme for Research and Innovation
through the Project No. 862644 (FET Open QUARTET).
S.P. acknowledges PRACE for awarding access to the Fenix Infrastructure resources at Cineca, which are partially funded by the European Union’s Horizon 2020 research and innovation program through the ICEI project under the Grant Agreement No. 800858. 
S. Cantori acknowledges partial support from the B-GREEN project of the italian MiSE - Bando 2018 Industria Sostenibile.
The authors acknowledge the use of IBM Quantum services for this work. The views expressed are those of the authors, and do not reflect the official policy or position of IBM or the IBM Quantum team.

\bibliographystyle{myunsrturl}
\bibliography{mybibliography}
 
\end{document}